\documentclass[a4paper,10pt]{article}
\usepackage[a4paper, margin=2cm]{geometry}

\usepackage[utf8]{inputenc}
\usepackage[T1]{fontenc}

\usepackage{fourier}

\usepackage{amsmath}
\usepackage{amssymb}
\allowdisplaybreaks
\usepackage{amsfonts}
\usepackage{bm}
\usepackage{mathrsfs}
\usepackage{booktabs}
\usepackage{graphicx}
\usepackage{upgreek}
\usepackage{color}
\usepackage[normalem]{ulem}
\makeatletter
\newcommand*{\transpose}{%
	{\mathpalette\@transpose{}}%
}
\newcommand*{\@transpose}[2]{%
	\raisebox{\depth}{$\m@th#1\intercal$}%
}
\makeatother

\usepackage{accents}
\newcommand{\ubar}[1]{\underaccent{\tilde}{#1}}

\usepackage{amsthm}

\makeatletter
\def\th@remark{%
	\thm@headfont{\bfseries}%
	\normalfont % body font
	\thm@preskip\topsep \divide\thm@preskip\tw@
	\thm@postskip\thm@preskip
}
\makeatother

\theoremstyle{remark}
\newtheorem*{remark}{Remark}

\usepackage[english]{babel}
\usepackage{csquotes}

\usepackage[hidelinks]{hyperref}

\title{Acoustoelastic analysis of soft viscoelastic  solids\\ with application to pre-stressed phononic crystals}
\author{Harold Berjamin \textsuperscript{a}, Riccardo De Pascalis \textsuperscript{b}\\
	{\footnotesize
		\begin{tabular}{l}
			~ \\
			\textsuperscript{a}School of Mathematical and Statistical Sciences, NUI Galway, University Road, Galway, Republic of Ireland \\
			\textsuperscript{b}Dipartimento di Matematica e Fisica \lq E. De Giorgi\rq , Universit\`a del Salento, Via per Arnesano, 73100, Lecce, Italy
	\end{tabular}}
}
\date{}

\usepackage{natbib}
\begin{document}
	
\maketitle
	
\begin{abstract}
	\noindent
	The effective dynamic properties of specific periodic structures involving rubber-like materials can be adjusted by pre-strain, thus facilitating the design of custom acoustic filters. While nonlinear viscoelastic behaviour is one of the main features of soft solids, it has been rarely incorporated in the study of such phononic media. Here, we study the dynamic response of nonlinear viscoelastic solids within a `small-on-large' acoustoelasticity framework, that is we consider the propagation of small amplitude waves superimposed on a large static deformation. Incompressible soft solids whose behaviour is described by the Fung--Simo quasi-linear viscoelasticy theory (QLV) are considered.  We derive the incremental equations using stress-like memory variables governed by linear evolution equations. Thus, we show that wave dispersion follows a strain-dependent generalised Maxwell rheology. Illustrations cover the propagation of plane waves under homogeneous tensile strain in a QLV Mooney--Rivlin solid. The acoustoelasticity theory is then applied to phononic crystals involving a lattice of hollow cylinders, by making use of a dedicated perturbation approach. In particular, results highlight the influence of viscoelastic dissipation on the location of the first band gap. We show that dissipation shifts the band gap frequencies, simultaneously increasing the band gap width. These results are relevant to practical applications of soft viscoelastic solids subject to static pre-stress.
	
	\medskip
	
\end{abstract}

%\tableofcontents

\medskip

\section{Introduction}\label{sec:Intro}

Solid rubber has been used in various cultural and engineering applications since ancient Mesoamerican times \citep{hosler99}. Mechanically, elastomers (including rubber vulcanisates) are very soft solids, a property that is also found in many soft biological materials such as skin, blood vessels, muscle, lung or brain tissue \citep{holzapfel2014biomechanics,al2018biomechanics}. Due to their high strength, they can support very large elastic deformations. Moreover, they can exhibit large hysteresis loops in loading-unloading experiments, as well as creep and relaxation phenomena \citep{ciambella10}. From these observations, one deduces that the mechanical stress is not solely function of the deformation. In isothermal or isentropic configurations, it is therefore quite natural to consider finite-strain viscoelastic material models to account for the dissipation of mechanical energy.

Among various theories found in the modelling literature, a nearly-incompressible viscoelastic model with internal variables was introduced by \citet{simo87}. In the limit of perfect incompressibility, the latter amounts to Fung's quasi-linear viscoelasticity (QLV) \citep{fung93,depascalis14} when the corresponding relaxation function is a scalar Prony series (under these assumptions, we will refer to it as \lq Fung--Simo\rq\ model). Despite experiments revealing the limits of this modelling approach \citep{ciambella10}, it has remained a very popular theory due to its simplicity and its ability to reproduce the main features of nonlinear viscoelastic behaviour, see \citet{dpr18b,jridi19,helisaz2021} to name a few recent experimental studies.

`Small-on-large' incremental motions are obtained by superposition of an infinitesimal deformation on a large static deformation. Such deformations have lead to numerous studies in mechanics, e.g. with applications to the propagation of small-amplitude waves in pre-stressed composites and phononic crystals \citep{Galich-Rudykh2017,barnwell16}. As far as viscoelastic solids are concerned, the literature is less abundant. \citet{destrade09} studied incremental motions for soft solids of the differential type (i.e., nonlinear Kelvin--Voigt solids), which are unable to capture stress relaxation phenomena \citep{banks11}. \citet{parnell19} overcome this limitation by addressing QLV acoustoelasticity in a specific configuration. The present study extends QLV acoustoelasticity to more general settings by making use of stress-like memory variables that arise naturally in the expression of the stress. This way, we provide a guide to the analysis of incremental motions in incompressible Fung--Simo solids.

The phononic crystal described by \citet{barnwell16} consists of soft cylinders inserted periodically into a host material, thus forming a square lattice. For this specific system, recent works have shown how an applied pre-stress can influence wave propagation properties in the absence of dissipation \citep{depascalis2020}, including the control of the band gaps that arise in the dispersion diagrams. To design optimal structures for specific purposes (such as noise filtering and wave guiding, cf. \citet{deymier2013acoustic}), it is therefore essential to understand how the geometric and constitutive parameters can influence the band structure.

Despite these advances, the authors are not currently aware of a practical realisation of the system introduced by \citet{barnwell16}, or any variant. Nevertheless, this specific phononic crystal remains an interesting academic example of a mechanical system whose effective dynamic properties can be tuned by pre-stress. Although rubber-like materials are often assumed elastic, viscoelastic behaviour eventually introduces frequency-dependent dissipation. In the perspective of potential experimental realisations (e.g., based on polymer 3D printing), it is therefore interesting to analyse the effects of dissipation on the dispersion diagrams.

As far as material dissipation is concerned, the phononic crystals literature includes studies of spring-mass systems with damping and of continuous systems with linear viscoelastic behaviour \citep{hussein2009, hussein2010}. In particular, similar configurations to \citet{barnwell16} were investigated in the limit of linear viscoelasticity \citep{wang15,krushynska16}. Nevertheless, to the authors' present knowledge, continuous systems with nonlinear viscoelastic behaviour were seldom considered \citep{parnell19}.

In the present study, we investigate the effects of viscoelastic dissipation on the above-mentioned pre-stressed periodic structure from \citet{barnwell16} at low frequencies, i.e. about the first band gap. Using the plane-wave expansion method, the `small-on-large' QLV theory is used to analyse the propagation of viscoelastic Bloch waves in the phononic crystal. For this purpose, a dedicated viscoelastic perturbation is introduced, based on the assumption that the viscoelastic material parameters do not vary in space (contrary to the elastic parameters). We find that dissipation shifts the band gap frequencies, simultaneously increasing the band gap width. It is therefore crucial to account for viscoelastic behaviour in soft phononic media where dissipation produces non-negligible effects.

The paper is organised as follows. Section~\ref{sec:GovEq} presents the equations of motion, their incremental counterpart, as well as the propagation of plane waves in homogeneous pre-stressed media with Mooney--Rivlin QLV behaviour. In Section~\ref{sec:Phononic}, the theory is applied to the periodic structure of \citet{barnwell16} by considering the propagation of small-amplitude antiplane waves in the pre-stressed phononic crystal. The effect of loss is analysed by means of a perturbation approach. Conclusions and prospects are detailed in Section~\ref{sec:Conclusion}. In the Appendix~\ref{app:thermo}, we show how the thermodynamic analysis of \citet{berjamin20} can be adapted to the case of incompressible QLV solids.

\section{Acoustoelastic motion of viscoelastic solids}\label{sec:GovEq}

\subsection{Preliminaries}\label{ssec:GovEqGen}

In what follows, we present the basic equations of incompressible Lagrangian solid dynamics \citep{holzapfel00}. We consider a homogeneous and isotropic solid continuum on which no external volume force is applied. Furthermore, self-gravitation is neglected. A particle initially located at some position $\bm{X}$ of the reference configuration moves to a position $\bm{x}$ of the current configuration. The deformation gradient tensor is the second-order tensor defined as
\begin{equation}
	\bm F = \frac{\partial{\bm{x}}}{\partial{\bm{X}}} = \bm{I} + \text{Grad}\, \bm{u} \, ,
	\label{F}
\end{equation}
where $\bm{u} = \bm{x} - \bm{X}$ is the displacement field, $\bm{I}$ is the metric tensor, and $\text{Grad}$ denotes the gradient operator with respect to the material coordinates ${\bm{X}}$ (Lagrangian gradient). If the Euclidean space is described by an orthonormal basis $\lbrace \bm{e}_1, \bm{e}_2, \bm{e}_3 \rbrace$ and a Cartesian coordinate system, then $\bm I$ has Kronecker delta components $\bm{I} = [\delta_{ij}]$.

In the analysis, we consider \emph{incompressible} materials, for which the constraint of no volume dilatation
\begin{equation}
	J = \det\bm{F}  \equiv 1
	\label{Incomp}
\end{equation}
is prescribed at all times, so that the mass density $\rho$ is constant. Various strain tensors are defined as functions of $\bm F$, such as the left Cauchy--Green tensor $\bm{B} = \bm{F}\bm{F}^\transpose\!$, the right Cauchy--Green tensor $\bm{C} = \bm{F}^\transpose \bm{F}$, and the Green--Lagrange tensor $\bm{E} = \frac{1}{2} (\bm{C} - \bm{I})$. Sometimes, the principal stretches $\lambda_i$ are introduced. Their squares $\lambda_i^2$ correspond to the eigenvalues of $\bm{A} \in \left\lbrace\bm{B}, \bm{C}\right\rbrace$. Thus, the invariants $I_i$ of $\bm A$ are given by
\begin{equation}
	\begin{aligned}
		& I_1 = \text{tr}\, \bm{A} = \lambda_1^2 + \lambda_2^2 + \lambda_3^2 \\
		& I_2 = \tfrac12 \big((\text{tr}\, \bm{A})^2 - \text{tr} (\bm{A}^2)\big) = \lambda_1^{-2} + \lambda_2^{-2} + \lambda_3^{-2} \\
		& I_3 = \det \bm{A} = \lambda_1^2 \lambda_2^2 \lambda_3^2 \equiv 1 \, ,
	\end{aligned}
	\label{Invariants}
\end{equation}
under the incompressibility constraint \eqref{Incomp}. Note that $I_3$ is related to the volume dilatation \eqref{Incomp} through $I_3 = J^2$.

In the absence of body forces, the motion is also governed by the conservation of momentum
\begin{equation}
	\rho \dot{\bm v} = \text{Div}\, \bm{P}
	\qquad\text{or}\qquad
	\rho \dot{\bm v} = \text{div}\, \bm{T}
	\label{EqMot}
\end{equation}
where the first Piola--Kirchhoff stress tensor $\bm P$ and the Cauchy stress tensor $\bm{T} = \bm{P} \bm{F}^\transpose = \bm{T}^\transpose$ are specified by the constitutive law. These stress tensors are also related to the second Piola--Kirchhoff stress tensor $\bm{S} = \bm{F}^{-1} \bm{P}$ which satisfies $\bm{T} = \bm{F}\bm{S}\bm{F}^\transpose$ for incompressible solids. While \lq$\text{Div}$\rq\ is the Lagrangian gradient's trace, the differential operator \lq$\text{div}$\rq\ is computed with respect to $\bm x$ (Eulerian divergence). The \lq dot\rq\ denotes the material time derivative.

The present definitions are consistent with notation used in the monograph by \citet{holzapfel00}. The divergence in Eq.~\eqref{EqMot} reads $[\text{div}\, \bm{T}]_i = T_{ij,j}$ componentwise, where indices after the coma denote spatial differentiation, and summation over repeated indices is performed. In some other texts, a transposed definition of the divergence is used, see e.g. \citet{destrade09}. In this case, the Lagrangian equation of motion involves the material divergence of the nominal stress tensor $\bm{P}^\transpose\!$ instead of $\bm P$.

\subsection{Fung--Simo incompressible viscoelasticity}\label{ssec:FungModel}

Fung's \emph{quasi-linear viscoelasticity} (QLV) or Fung's model of viscoelasticity is presented below (see Section~7.13 of \citet{fung93}). This model is based on the assumption that the stress is linearly dependent on the history of the elastic stress response  by considering   a fading memory effect   with a Boltzmann superposition principle  and  that the viscous relaxation rate is independent of the instantaneous local strain. By analogy with linear viscoelasticity \citep{carcione15}, the second Piola--Kirchhoff stress is therefore given by \citep{fung93}
\begin{equation}
	\bm{S} = {\mathbb G} \bm{*} \dot{\bm S}^\text{e} = \int_{-\infty}^{+\infty} {\mathbb G}(t-s) : \dot{\bm S}^\text{e}(s) \, \text d s = \dot{{\mathbb G}} \bm{*} {\bm S}^\text{e} 
	\label{FungConv}
\end{equation}
for compressible solids, where the colon denotes double contraction ${\mathbb G} : \dot{\bm S}^\text{e} = [{G}_{ijk\ell} {\dot{S}^\text{e}}_{k\ell}]$. The stress tensor $\bm{S}^\text{e} = 2\, \partial W/\partial\bm{C}$ is the elastic response derived from a strain energy density function $W$, and ${\mathbb G}$ is a fourth-order relaxation tensor.

Although other choices can be made, scalar relaxation is sometimes assumed \citep{taylor09, wineman09} by choosing the fourth-order relaxation tensor ${\mathbb G} = \mathscr{G}\, {\mathbb I}^\text{s}$ along the symmetric identity tensor ${\mathbb I}^\text{s} = \frac{1}{2} [\delta_{ik}\delta_{j\ell} + \delta_{i\ell}\delta_{jk}]$. The relaxation function $\mathscr{G}$ is assumed proportional to the Heaviside step function $\operatorname{H}$ (i.e., $\mathscr{G}$ is of the Heaviside type, see \citet{carcione15}). Typically, the relaxation function $\mathscr{G}$ may be chosen as a Prony series of the form
\begin{equation}
	\mathscr{G}(t) = \bigg[1 - \sum_{k=1}^n g_k\, (1-\text{e}^{-t/\tau_k})\bigg] \operatorname{H}(t) \, ,
	\label{PronyRel}
\end{equation}
with an arbitrary number $n$ of relaxation mechanisms of magnitude $g_k$ and characteristic time $\tau_k$.

For causal problems where deformation starts at $t=0$, the elastic response $\bm{S}^\text{e}$ is of the Heaviside type too. Thus, the product rule of differentiation and restriction of the integrals to $[0,t]$ yields alternative forms of the convolution products \citep{fung93,depascalis14}. In the present study, the material is not assumed stress-free at negative times. We thus keep convolution over $\mathbb R$ instead, see Eq.~\eqref{FungConv}.

In the incompressible case \eqref{Incomp}, the stress response $\bm{S}$ includes an additional term $-p\bm{C}^{-1}$ where $p$ is the Lagrange multiplier of the incompressibility constraint{\,---\,}the corresponding Cauchy stress ${\bm T} = \bm{F}{\bm S}{\bm F}^\transpose\!$ includes the term $-p\bm{I}$. Moreover, the strain energy function $W$ depends on the invariants $I_1$, $I_2$ only, and the elastic response reduces to
\begin{equation}
	\bm{S}^\text{e} = 2 \left(W_1 + I_1 W_2\right) \bm{I} - 2W_2 \bm{C} \, ,
	\label{Elast}
\end{equation}
where $W_i$ is shorthand for the partial derivative $\partial W/\partial I_i$ evaluated at $(I_1, I_2)$. Equivalent forms to Eq.~\eqref{Elast} can be derived by using the Cayley--Hamilton theorem, but we do not enter into these considerations here (see \citet{depascalis14}). For sake of consistency with incompressible linear elasticity in the limit of infinitesimal deformations, we assume that the relationship $\mu/2 = W_1(3,3) + W_2(3,3)$ defines the shear modulus $\mu > 0$.

To split stresses into deviatoric/isochoric and hydrostatic/volumetric contributions, \citet{simo87} introduced a nearly incompressible theory with internal variables. Integration of the differential equations governing the evolution of internal variables allows us to rewrite the constitutive law using hereditary integrals. In the incompressible limit, we thus have (see Eq.~(1.17) of \citet{simo87})
\begin{equation}
	\begin{aligned}
		\bm{S} &= -p\bm{C}^{-1}\! + \text{Dev}\left(\mathscr{G} * \dot{\bm S}^\text{e}_\text{D}\right) \\
		&= -q\bm{C}^{-1}\! + \mathscr{G} * \dot{\bm S}^\text{e}_\text{D} = -q\bm{C}^{-1}\! + \int_{\mathbb R} \mathscr{G}(t-s) \dot{\bm S}^\text{e}_\text{D}(s)\, \text{d}s = -q\bm{C}^{-1}\! + \dot{\mathscr{G}} * {\bm S}^\text{e}_\text{D}
	\end{aligned}
	\label{FungConvSimo}
\end{equation}
with $\mathscr{G}$ defined in Eq.~\eqref{PronyRel}, up to a suitable redefinition of the arbitrary Lagrange multiplier $p$ as $q$. Here, we have introduced the notation ${\bm S}^\text{e}_\text{D} = \text{Dev}(\bm{S}^\text{e})$ where $\text{Dev}(\bullet) = (\bullet) - \tfrac13 (\bullet :\bm{C}) \bm{C}^{-1}$ denotes the \emph{deviatoric operator} in the Lagrangian description \citep{holzapfel00}. Similarly to Eq.~\eqref{FungConv}, the star operator $*$ denotes the standard convolution product in time domain, and the elastic response is deduced from Eq.~\eqref{Elast}.

Let us assume that the material is in an equilibrium state $\bm{S}^\text{e} = \bar{\bm S}^\text{e}$ for negative times, which undergoes continuous perturbations about $t=0$. The rate of ${\bm S}^\text{e}_\text{D}$ is therefore a causal signal which vanishes at the origin of times. The expression of the relaxation function \eqref{PronyRel} in Eq.~\eqref{FungConvSimo} yields
\begin{equation}
	\bm{S} = -q \bm{C}^{-1}\! + {\bm S}^\text{e}_\text{D} - \sum_{k=1}^n \bm{S}^\text{v}_k \, ,
	\qquad
	\bm{T} = -q\bm{I} + {\bm T}^\text{e}_\text{d} - \sum_{k=1}^n \bm{T}^\text{v}_k \, ,
	\label{FungMemStress}
\end{equation}
where
\begin{equation}
	\bm{S}^\text{v}_k = g_k \int_0^t \big(1-\text{e}^{-(t-s)/\tau_k}\big)\, \dot {\bm S}^\text{e}_\text{D}(s) \text d s = \frac{g_k}{\tau_k} \int_0^t \text{e}^{-(t-s)/\tau_k}\, {\bm S}^\text{e}_\text{D}(s) \text d s 
	\label{FungMemVar}
\end{equation}
and $\bm{T}^\text{v}_k = \bm{F}\bm{S}^\text{v}_k\bm{F}^\transpose$
are \emph{memory variables} arising in the expression of the convolution product \citep{berjamin20}. Here, we have introduced the notation ${\bm T}^\text{e}_\text{d} = \text{dev}(\bm{T}^\text{e})$, where the operator $\text{dev}(\bullet) = (\bullet) - \frac13 \text{tr}(\bullet) \bm{I}$ is the deviatoric projection in the spatial description \citep{holzapfel00}. Now, computing the time derivative of the memory variables, we find that $\bm{S}^\text{v}_k$ satisfies the linear evolution equation \citep{taylor09}
\begin{equation}
	\tau_k \dot{\bm S}^\text{v}_k = g_k {\bm S}^\text{e}_\text{D} - \bm{S}^\text{v}_k\, .
	\label{FungEvol}
\end{equation}
Thus, the convolution product in the constitutive law \eqref{FungConvSimo} is replaced by a summation of $n$ memory variables which satisfy a linear differential equation. This way, the Fung--Simo QLV theory introduces an additive decomposition of stress \citep{berjamin20}. One notes that this model is equivalent to the incompressible version of the Fung-type viscoelastic model described by \citet{depascalis14}, as noted in other related works \citep{balbi18, berjamin20}. Thermodynamic consistency is discussed in Appendix~\ref{app:thermo}.

\subsection{Small-on-large analysis}\label{sec:Incr}

\begin{figure}
	\centering
	\includegraphics{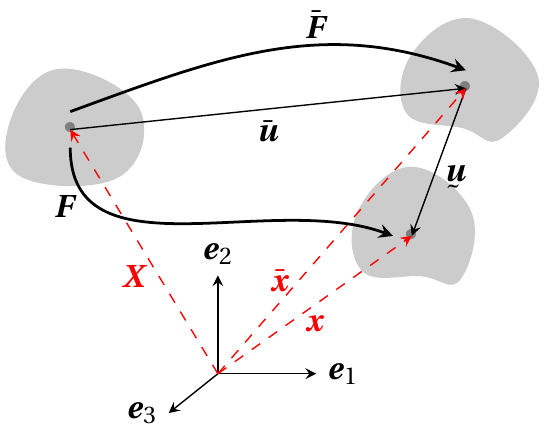}
	\caption{Acoustoelasticity. Combination of a large static deformation and a small incremental perturbation. \label{fig:Config}}
\end{figure}

For the derivation of the QLV acoustoelasticity equations, we follow the same steps as in \citet{destrade09}, referring to Section~\ref{ssec:GovEqGen} for definitions and notations. A similar derivation is presented in \citet{parnell19} without the use of memory variables. The main idea consists in introducing an intermediate equilibrium configuration such that the total finite motion results from an infinitesimal perturbation of the former. This approach is commonly referred to as \lq incremental\rq\ or \lq small on large\rq\ theory \citep{ogden1997non}. As shown in Figure~\ref{fig:Config}, the solid undergoes a finite static pre-deformation $\bm{X} \mapsto \bar{\bm x}(\bm{X})$ followed by a dynamic infinitesimal deformation $\bar{\bm x}(\bm{X}) \mapsto {\bm x}(\bar{\bm x}(\bm{X}),t)$, which is pushed forward from the reference configuration into a new one through the intermediate state.

\paragraph{Static deformation}

At equilibrium, the deformation is governed by the balance equation \eqref{EqMot} with zero velocity $\bar{\bm v} = \bm{0}$. We denote all quantities associated with the present finite deformation by an overbar. Thus, the corresponding displacement vector is defined by $\bar{\bm u} = \bar{\bm x} - \bm{X}$. The stresses are deduced from the deformation gradients $\bar{\bm F} = \partial\bar{\bm x}/\partial\bm{X}$ and from the constitutive law \eqref{FungMemStress}, where the memory variables \eqref{FungEvol} equal their equilibrium value $\bar{\bm S}_k^\text{v} = g_k \bar{\bm S}^\text{e}_\text{D}$, or equivalently $\bar{\bm T}_k^\text{v} = g_k \bar{\bm T}^\text{e}_\text{d}$. Thus, the material is in its relaxed elastic limit, which satisfies
\begin{equation}
	\overline{\text{div}}\, \bar{\bm T} = \bm{0}
	\qquad\text{with}\qquad
	\bar{\bm T} = -\bar{q} {\bm I} + \left(1 - \sum_{k=1}^n g_k \right) \bar{\bm T}^\text{e}_\text{d}
	\label{StaticStress}
\end{equation}
deduced from the constitutive law \eqref{FungMemStress}, or equivalently ${\text{Div}}\, \bar{\bm P} = \bm{0}$ with the stress $\bar{\bm P} = \bar{\bm T} \bar{\bm F}^{-\transpose}\!$. The divergence operator with the overbar is obtained by spatial differentiation with respect to $\bar{\bm x}$. We will see later on how it relates to $\text{div}$.

\paragraph{Dynamic perturbation}

Now, a deformation with small displacement $\ubar{\bm u}$ is superimposed on the present finite static deformation. We denote incremental quantities associated with the infinitesimal deformation by an undertilde. For instance, the total displacement and position fields are respectively given by $\bm{u} = \bar{\bm u} + \ubar{\bm u}$ and $\bm{x} = \bar{\bm x} + \ubar{\bm u}$, where the explicit dependence on spatial and temporal coordinates has been omitted for simplicity (see Fig.~\ref{fig:Config}). Using the chain rule, the deformation gradient tensor is decomposed as
\begin{equation}
	\bm{F} = \frac{\partial \bm{x}}{\partial \bar{\bm x}} \frac{\partial \bar{\bm x}}{\partial \bm{X}} = (\bm{I} + \bm{H}) \bar{\bm F} = \bar{\bm F} + \ubar{\bm F} \, ,\qquad \ubar{\bm F} = {\bm H} \bar{\bm F} \, ,
	\label{FDec}
\end{equation}
where ${\bm H} = {\partial \ubar{\bm u}}/{\partial \bar{\bm x}}$ is the incremental \emph{displacement gradient} tensor in the pre-deformed configuration (the undertilde is discarded for this incremental quantity for sake of parsimony, as there is no ambiguity). Introducing the infinitesimal strain tensor $\bm\varepsilon$ such that $2{\bm \varepsilon} = {\bm H} + {\bm H}^\transpose\!$, we compute the incremental strain tensor $\ubar{\bm C} = \bar{\bm F}^\transpose (2 {\bm \varepsilon}) \bar{\bm F}$, as well as the incremental invariants \citep{destrade09}
\begin{equation}
	\ubar I_{1} = 2\, \text{tr}( {\bm \varepsilon} \bar{\bm B}) \, , \qquad
	\ubar I_{2} = -2\, \text{tr}\big( {\bm \varepsilon} \bar{\bm B}^{-1}\big) \, , \qquad
	\ubar I_{3} = 0 \, ,
\end{equation}
where the expression for $\ubar I_{2}$ is deduced from the Cayley--Hamilton identity applied to the tensor $\bar{\bm B}$.

Let us write the Eulerian equation of motion \eqref{EqMot} corresponding to the total deformation, with the displacement field ${\bm u}$ and the Cauchy stress tensor ${\bm T} \simeq \bar{\bm T} + \ubar{\bm T}$. Since the displacement $\bar{\bm u}$ is static, the total velocity $\bm{v} = \dot{\bm x}$ reduces to the material time-derivative of $\ubar{\bm u}$. One notes that the divergence operators are linked through
\begin{equation}
	\text{div}\, {\bm T} = \overline{\text{div}} \left({\bm T} (\bm{I}+\bm{H})^{-\transpose}\right) \simeq \overline{\text{div}} \left({\bm T} - {\bm T}\bm{H}^{\transpose}\right) ,
\end{equation}
where we have used the Piola identity over the incremental deformation $\bar{\bm x} \mapsto \bm{x}$ with deformation gradient tensor $\bm{I}+\bm{H}$ (see Eq.~\eqref{FDec}). Conversely, we note that the relationship $\overline{\text{div}}\, {\bm T} = \text{div}({\bm T} + {\bm T}\bm{H}^{\transpose})$ is satisfied at leading order too. These relationships between divergence operators are of the same type as Eq.~\eqref{EqMot} (see Sec.~4.3 of \citet{holzapfel00}), up to a redefinition of the reference and deformed states using the intermediate configuration of Fig.~\ref{fig:Config}.

Since the pre-deformed solid is in a static equilibrium characterised by Eq.~\eqref{StaticStress}, we end up with the incremental equation of motion
\begin{equation}
	\rho \ubar{\ddot{\bm u}} = \text{div}\, \ubar{\bm \Sigma} , \qquad \ubar{\bm \Sigma} = \ubar{\bm T} - \bar{\bm T} \bm{H}^\transpose  ,
	\label{IncrementMot}
\end{equation}
where the use of $\text{div}$ or of $\overline{\text{div}}$ is equivalent at the same order of approximation. In fact, the position vector ${\bm x} \simeq \bar{\bm x}$ describes a small perturbation from the pre-deformed state, which is reminiscent of the linear elastic framework.
The incremental incompressibility constraint deduced from Eq.~\eqref{Incomp} and from the decomposition \eqref{FDec} reads $\text{tr}\, {\bm H} = 0$, or equivalently $\text{div}\,\ubar{\bm u} = 0$.

Using the constitutive law \eqref{FungMemStress}, we derive the expression of the incremental Cauchy stress tensor as follows
\begin{equation}
	\ubar{\bm T} = -\ubar{q}\bm{I} + \ubar{\bm T}^\text{e}_\text{d} - \sum_{k=1}^n \ubar{\bm T}^\text{v}_k \, ,
	\label{IncrementCauchy}
\end{equation}
where $\ubar{q}$ is the incremental pressure. The increment of the elastic response is deduced from the definition ${\bm T}^\text{e} = \bm{F}{\bm S}^\text{e}\bm{F}^\transpose$ by linearising the products, and a similar computation is performed for the viscous stresses. Thus, the effective linearised stress of Eq.~\eqref{IncrementMot} reads
\begin{equation}
	\ubar{\bm \Sigma} = -\ubar{q}\bm{I} + \bar{q} \bm{H}^\transpose + \left(1-\sum_{k=1}^n g_k\right) {\bm H} \bar{\bm T}^\text{e}_\text{d} + \bar{\bm F}\left(\ubar{\bm S}^\text{e}_\text{D} - \sum_{k=1}^n \ubar{\bm S}_k^\text{v}\right)\bar{\bm F}^\transpose.
	\label{IncrementStress}
\end{equation}
The relevant increments deduced from Eq.~\eqref{Elast} are given by
\begin{equation}
	\ubar{\bm S}^\text{e} = 2 \left(\ubar{W}_1 + \ubar{I}_1 \bar{W}_2 + \bar{I}_1 \ubar{W}_2\right) \bm{I} - 2 \ubar{W}_2 \bar{\bm C} - 2\bar W_2 \ubar{\bm C}\, ,
	\label{IncrementConstitutive}
\end{equation}
and thus
\begin{equation}
	\begin{aligned}
		\ubar{\bm S}^\text{e}_\text{D} &= \ubar{\bm S}^\text{e} - \tfrac13 \left(\ubar{\bm S}^\text{e}:\bar{\bm C} + \bar{\bm S}^\text{e}:\ubar{\bm C}\right) \bar{\bm C}^{-1} - \tfrac13 \left(\bar{\bm S}^\text{e}:\bar{\bm C}\right) \ubar{\bm C}^{-1} \\
		&= \ubar{\bm S}^\text{e} - \tfrac23 \left(\ubar I_1\bar W_1 + 2 \ubar I_2 \bar W_2 + \bar I_1\ubar W_1 + 2 \bar I_2 \ubar W_2\right) \bar{\bm C}^{-1} - \tfrac23 \left(\bar I_1\bar W_1 + 2 \bar I_2 \bar W_2\right) \ubar{\bm C}^{-1} 
	\end{aligned}
	\label{IncrementDev}
\end{equation}
with $\ubar{\bm C}^{-1}  = -\bar{\bm F}^{-1} (2\bm{\varepsilon}) \bar{\bm F}^{-\transpose}$.
From Eq.~\eqref{FungEvol}, we deduce
\begin{equation}
	\tau_k\, \dot{\ubar{\bm S}}_k^\text{v} = g_k \ubar{\bm S}^\text{e}_\text{D} - \ubar{\bm S}_k^\text{v} \, .
	\label{IncrementEvol}
\end{equation}
In Eqs.~\eqref{IncrementConstitutive}-\eqref{IncrementDev}, the increment $\ubar{W}_i = W_i - \bar W_i$ given by $\ubar{W}_i = \ubar{I}_1\bar{W}_{i1}  + \ubar{I}_2\bar{W}_{i2}$ follows from a truncated Taylor series of the function $W_i(I_1, I_2)$ about the equilibrium values $(\bar I_1, \bar I_2)$, where $W_{ij}$ is shorthand for ${\partial W_i}/{\partial I_j}$, see \citet{destrade09}. If we compute the transpose of Eq.~\eqref{IncrementStress}, then we remark that the incremental stress tensor $\ubar{\bm \Sigma}$ is not preserved. Moreover, substitution of ${\bm H}$ by its transpose does not necessarily keep $\ubar{\bm \Sigma}$ invariant, showing that the incremental constitutive law \eqref{IncrementStress} does not exhibit any elementary symmetry in general.

\subsection{Dispersive plane waves}\label{sec:IncrDisp}

We consider incremental harmonic plane waves of the form
\begin{equation}
	\ubar{\bm u} = \hat{\bm u}\, \text{e}^{\text{i}(\omega t - \kappa\bm{n \cdot x})} \, , \qquad
	{\bm H} = \hat{\bm H}\, \text{e}^{\text{i}(\omega t - \kappa\bm{n \cdot x})} ,\qquad
	\ubar{q} = \hat{q}\, \text{e}^{\text{i}(\omega t - \kappa\bm{n \cdot x})} ,
	\label{Harmonic}
\end{equation}
with complex amplitude $\hat{\bm u}$ for the incremental displacement field $\ubar{\bm u}$, and similar notation is used for the harmonic amplitude of other incremental quantities such as $\hat q$, $\hat{\bm \Sigma}$, $\hat{\bm T}$, etc. Following from the definition of $\bm H$, the displacement gradient's harmonic amplitude satisfies $\hat{\bm H} = -\text{i} \kappa\, (\hat{\bm u}\otimes\bm{n})$. The exponential space-time dependency involves the angular frequency $\omega$, the wave number $\kappa$, and the imaginary unit $\text{i} = \sqrt{-1}$. This wave is propagating along the $\bm{n}$-direction, where $\bm{n}$ is an arbitrary unit vector.

Injecting this Ansatz in the incremental incompressibility constraint gives us the orthogonality condition $\hat{\bm u} \cdot \bm{n} = 0$, i.e. the wave corresponds to a transverse shearing motion. The incremental equation of motion reads $\rho\omega^2 \hat{\bm u} = \text{i}\kappa \hat{\bm \Sigma} \bm{n}$ in harmonic form, where the complex amplitudes 
\begin{equation}
	\hat{\bm \Sigma} = -\hat{q}\bm{I} + \bar{q} \hat{\bm H}^\transpose + \left(1-\sum_{k=1}^n g_k\right) \hat{\bm H} \bar{\bm T}^\text{e}_\text{d} + \left(1-\sum_{k=1}^n \frac{g_k}{1+\text{i}\omega\tau_k}\right) \bar{\bm F} \hat{\bm S}^\text{e}_\text{D} \bar{\bm F}^\transpose
	\label{IncrementStressDyn}
\end{equation}
are deduced from Eqs.~\eqref{IncrementStress}-\eqref{IncrementEvol}. Given the definition of $\bm H$, the displacement gradient amplitudes $\hat{\bm H} = -\text{i} \kappa\, (\hat{\bm u}\otimes\bm{n})$ satisfy the property $\hat{\bm H}^\transpose \bm{n} = \bm{0}$ due to the orthogonality condition. Then, scalar multiplication of the wave equation by the unit vector $\bm n$ yields the condition $\bm{n}^\transpose \hat{\bm \Sigma} \bm{n} = 0$ from which the dynamic pressure $\hat q$ is deduced, see expression of $\hat{\bm \Sigma}$ in Eq.~\eqref{IncrementStressDyn}.
Given Eq.~\eqref{IncrementStressDyn}, one observes that the incremental stress $\hat{\bm \Sigma}$ is linear in the displacement gradient amplitudes $\hat{\bm H}$. In other words, we can write a relationship of the form
\begin{equation}
	\hat{\bm \Sigma}\bm{n} = [\bm{I} - \bm{n}\otimes\bm{n}] \left[\hat{\bm \Sigma} + \hat{q}\bm{I} - \bar{q} \hat{\bm H}^\transpose\right] \bm{n}
	\qquad\text{where}\qquad
	\hat{\bm \Sigma} + \hat{q}\bm{I} - \bar{q} \hat{\bm H}^\transpose = {\mathbb A}:\hat{\bm H}, 
	\label{SoLStress}
\end{equation}
for some fourth-order \emph{instantaneous stiffness tensor} ${\mathbb A}$ to be determined in practical cases. In general, this tensor does not present any specific symmetry. The first equation of Eq.~\eqref{SoLStress} follows from the property $\bm{n}^\transpose \hat{\bm \Sigma} \bm{n} = 0$ where $\bm{n}$ is unitary.

Finally, let us introduce the second-order \emph{acoustic tensor} $\bm{Q}$ such that $[{\mathbb A}:\hat{\bm H}] \bm{n} = -\text i\kappa \bm{Q} \hat{\bm u}$, i.e. which components are given by $Q_{ij} = A_{ipjq} n_p n_q$. In general, this tensor is complex-valued and not necessarily symmetric. The incremental wave equation rewrites as an eigenvalue problem of the form
\begin{equation}
	\rho \frac{\omega^2}{\kappa^2} \hat{\bm u} = [\bm{I} - \bm{n}\otimes\bm{n}] \bm{Q} [\bm{I} - \bm{n}\otimes\bm{n}]\, \hat{\bm u}
	\label{SoLDisp}
\end{equation}
governing nontrivial solutions $\hat{\bm u}\neq {\bf 0}$, where we have used Eq.~\eqref{SoLStress} and the orthogonality condition $\hat{\bm u} \cdot \bm{n} = 0$ \citep{scott85}. The above dispersion relationship links the wave number $\kappa$ to the frequency $\omega$. Assuming that an admissible polarisation vector is known, the dynamic modulus $\rho {\omega^2}/{\kappa^2}$ is then deduced from Eq.~\eqref{SoLDisp} by scalar multiplication with $\hat{\bm u}$. Note in passing that \emph{frequency-deformation separability} is not satisfied for general plane waves, i.e. $\rho {\omega^2}/{\kappa^2}$ cannot necessarily be written as the product of one function of $\omega$ and one function of $\bar{\bm F}$. \citet{parnell19} found that this property was satisfied for elongated slender beams at long times.

\begin{remark}
	If no pre-deformation is applied (i.e., if $\bar{\bm F} = \bm{I}$ and $\bar{\bm T} = \bm{0}$), we note that the incremental stresses satisfy $\ubar{\bm \Sigma} = \ubar{\bm T}$, see Eq.~\eqref{IncrementMot}, and that the elastic stress $\bar{\bm T}^\text{e}_\text{d}$ deduced from Eq.~\eqref{Elast} with $\bar I_1 = 3$ vanishes too. Using Eqs.~\eqref{IncrementConstitutive}-\eqref{IncrementDev}, the expression in Eq.~\eqref{IncrementStressDyn} reduces to
	\begin{equation}
		\hat{\bm \Sigma} = -\hat{q}^\star \bm{I} + 2 \mu \left(1-\sum_{k=1}^n \frac{g_k}{1+\text{i}\omega\tau_k}\right) \hat{\bm \varepsilon}
		\label{StressLin}
	\end{equation}
	in the absence of pre-deformation, up to an appropriate redefinition of the dynamic pressure $\hat q$ as $\hat q^\star\!$. Here, we have used the fact that the elastic response is consistent with linear elasticity in the infinitesimal strain limit. One observes that a generalised Maxwell rheology is recovered, as expected.
\end{remark}

\subsection{Illustration}\label{subsec:Illustr}

For illustration purposes, let us consider the propagation of transverse shear waves in an unbounded material undergoing simple tension/compression in the propagation direction. The material is assumed to have incompressible Mooney--Rivlin behaviour in the elastic range, i.e. the strain energy function reads
\begin{equation}
	W = \mathfrak{C}_1 \left(I_1 - 3\right) + \mathfrak{C}_2 \left(I_2 - 3\right)
	\label{WMooney}
\end{equation}
where the invariants are defined in Eq.~\eqref{Invariants}. For consistency with linear elasticity, the material parameters $\mathfrak{C}_1 = W_1$ and $\mathfrak{C}_2 = W_2$ are constants related to the shear modulus $\mu = 2\left(\mathfrak{C}_1 + \mathfrak{C}_2\right)$. The neo-Hookean strain energy function is recovered when $\mathfrak{C}_2$ is equal to zero. Viscoelastic behaviour with $n=1$ relaxation mechanism is assumed, with parameters $\tau=\tau_1$ and $g = g_1$ for sake of simplicity.

Physical parameters for soft rubber-like solids can be estimated from various literature sources, see summary in Table~\ref{tab:Params}. For the long-time elastic response, we use the material parameters for incompressible Mooney--Rivlin rubber from \citet{marckmann06} extracted from the data by Treloar, i.e. $(1-g) \mathfrak{C}_1 = 0.162$~MPa and $(1-g) \mathfrak{C}_2 = 5.9$~kPa.
Therefore, the relaxed shear modulus equals $0.336$~MPa. Comparable results were obtained by \citet{khajehsaeid13}, with the numerical values 0.142 and 0.011~MPa for the two long-time Mooney parameters.

The viscoelastic parameters $g$, $\tau$ are inferred from \citet{ciambella10}, where we have selected the first relaxation mechanism for the Fung model identified through a relaxation test. Thus, the unrelaxed shear modulus equals $\mu = 0.473$~MPa. In upcoming computations, the mass density $\rho = 1.1\times 10^3$ kg/m\textsuperscript{3} of soft rubber is assumed. Therefore, the shear wave speed $c = \sqrt{\mu/\rho}$ takes the numerical value $c \approx 21$~m/s with the values of Table~\ref{tab:Params}.

\begin{table}
	\centering
	\caption{Reference values of the material parameters describing a rubber-like solid. The viscoelastic Mooney--Rivlin parameters for $n=1$ relaxation mechanism are extracted from several literature references \citep{marckmann06, ciambella10}. \label{tab:Params}}
	
	\vspace{0.5em}

	{\renewcommand{\arraystretch}{1.2}
		\begin{tabular}{ccccc}
			\toprule 
			$\rho$ [kg/m\textsuperscript{3}] & $\mathfrak{C}_1$ [MPa] & $\mathfrak{C}_2$ [kPa] & $g$ [---] & $\tau$ [s]  \\  \midrule  
			$1.1\times 10^3$ & $0.228$ & $8.3$ & $0.29$ & $0.31$ \\ 
			%\textsc{Neo--Hookean}   & $1.1\times 10^3$ & $0.2$ & $0$ & $0.29$ &$0.31$ \\
			\bottomrule
	\end{tabular}}
\end{table}

Now, consider that the material is deformed according to uniaxial tension-compression along $\bm{e}_3$, i.e. the deformation gradient reads $\bar{\bm F} = \text{diag}\left[\bar \lambda_i\right]$, with the axial stretch $\bar\lambda_3 = \bar\lambda$ and lateral stretches $\bar\lambda_1 = \bar\lambda_2$ such that incompressibility $\bar\lambda\bar\lambda_1^2 = 1$ is enforced. The equilibrium equation \eqref{StaticStress} must be satisfied, where the stress $\bar{\bm T}$ follows from the elastic response $\bar{\bm T}^\text{e} = \bar{\bm F}\bar{\bm S}^\text{e}\bar{\bm F}^\transpose\!$ of Eq.~\eqref{Elast} with the invariants given by $\bar I_1 = \bar \lambda^2 + 2/\bar \lambda$ and $\bar I_2 = \bar \lambda^{-2} + 2\bar \lambda$. Thus, the equilibrium stress is a diagonal tensor whose entries are the principal stresses, and the lateral tractions are equal. Imposing that the lateral tractions $\bar{T}_{11}=\bar{T}_{22}$ vanish leads to the expression of the Lagrange multiplier $\bar q$ and of the axial equilibrium stress $\bar{T}_{33}$.

Incremental wave solutions propagating in a direction $\bm{n}$ normal to the direction of elongation $\bm{e}_3$ are considered, for instance $\bm{n} = \bm{e}_1$. In agreement with the orthogonality condition, we assume that the wave is polarised along $\bm{e}_3$, so that the displacement gradient tensor $\hat{\bm H} = -\text{i} \kappa\, (\hat{\bm u}\otimes\bm{n})$ has nonzero components along $\bm{e}_3\otimes \bm{e}_1$ only. Since the pre-deformation is homogeneous \citep{destrade09}, the incremental wave equation \eqref{IncrementMot} becomes $\rho\omega^2 \hat{\bm u} = \text{i}\kappa \hat{\bm T} \bm{n}$ in harmonic form. Following the same steps as in the above derivation, the harmonic amplitude of the incremental Cauchy stress \eqref{IncrementCauchy} is obtained:
\begin{equation}
	\hat{\bm T} = -\hat{q}\bm{I}  + (1 - g) \left(\hat{\bm H} \bar{\bm T}^\text{e}_\text{d} + \bar{\bm T}^\text{e}_\text{d} \hat{\bm H}^\transpose\right) + \left(1 - \frac{g}{1+\text{i}\omega\tau}\right) \bar{\bm F} \hat{\bm S}^\text{e}_\text{D} \bar{\bm F}^\transpose .
\end{equation}
Using Eqs.~\eqref{IncrementConstitutive}-\eqref{IncrementDev} and the fact that $\bar{\bm T}^\text{e}_\text{d}$ is diagonal, multiplication by $\bm n$ then yields
\begin{equation}
	\left[\hat{\bm T} + \hat q^\star \bm{I}\right]\bm{n} = \left[{\mathbb A}:\hat{\bm H}\right] \bm{n} = -\text{i}\kappa \bm{Q} \hat{\bm u} = -\text{i} \kappa \mu_x \hat{\bm u}
\end{equation}
up to a redefinition of the acoustic pressure $\hat q$, with the coefficients
\begin{equation}
	\mu_x = (1 - g) [\bar{T}^\text{e}_\text{d}]_{11} + \left(1 - \frac{g}{1+\text{i}\omega\tau}\right) \bar\mu_x^\text{v}
	\label{Dispersion}
\end{equation}
and
\begin{equation}
	[\bar{T}^\text{e}_\text{d}]_{11} = \frac23 \left[\left(3\bar\lambda^{-1} - \bar I_1\right) \mathfrak{C}_1 + \left(3\bar I_1 \bar\lambda^{-1} - 3 \bar\lambda^{-2} - 2 \bar I_2\right) \mathfrak{C}_2\right] ,\qquad
	\bar\mu_x^\text{v} = \frac23 \left[\bar{I}_1\mathfrak{C}_1 + \left(2\bar{I}_2 - 3\bar\lambda\right)\mathfrak{C}_2\right] .
\end{equation}
Here we have also used the expression of the left Cauchy--Green strain tensor $\bar{\bm B} = \text{diag}\left[\bar \lambda_i^2\right]$, and the property $2\hat{\bm \varepsilon} \bm{n} = -\text{i}\kappa\hat{\bm u}$ following from the orthogonality condition. Therefore, the instantaneous stiffness tensor ${\mathbb A} = 2 \mu_x {\mathbb I}^\text{s}$ is proportional to the fourth-order symmetric identity tensor, and the acoustic tensor ${\bm Q} = \mu_x {\bm I}$ is proportional to identity. Finally, the dispersion relationship $\rho {\omega^2}/{\kappa^2} = \mu_x$ for incremental displacements $\hat{\bm u}$ polarised along $\bm{e}_3$ is obtained.

\begin{figure}
	\centering
	\includegraphics{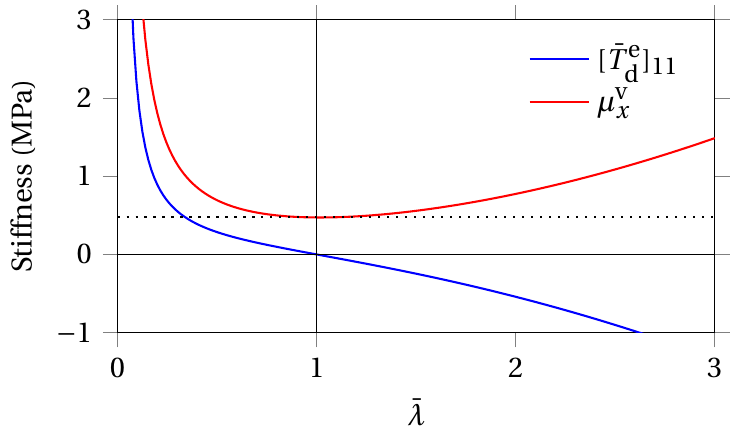}
	
	\caption{Acoustoelasticity of transverse waves propagating in the transverse direction to tensile stretch. Evolution of the coefficients involved in the expression \eqref{Dispersion} of the dynamic modulus $\mu_x$ in terms of the static stretch $\bar \lambda$. The horizontal dots mark the value of the shear modulus $\mu$. \label{fig:Acoustoelast}}
\end{figure}

Figure~\ref{fig:Acoustoelast} represents the evolution of the stiffness coefficients governing the two terms of Eq.~\eqref{Dispersion} with respect to the pre-stretch $\bar\lambda$. If no pre-deformation is applied ($\bar \lambda = 1$), then the stress component $[\bar{T}^\text{e}_\text{d}]_{11}$ vanishes, and we recover the same dynamic modulus $\rho {\omega^2}/{\kappa^2}$ as deduced from Eq.~\eqref{StressLin}. In the vicinity of the undeformed state, the coefficient $[\bar{T}^\text{e}_\text{d}]_{11}$ is decreasing with respect to $\bar\lambda$, whereas $\mu_x^\text{v}$ has reached a local minimum (zero slope). Note in passing that frequency-deformation separability is not satisfied in general, i.e. the dynamic modulus $\mu_x$ cannot be written as the product of one function of $\omega$ and one function of $\bar\lambda$.

\begin{figure}
	\begin{minipage}{0.49\textwidth}
		\centering
		(a)
		
		\includegraphics{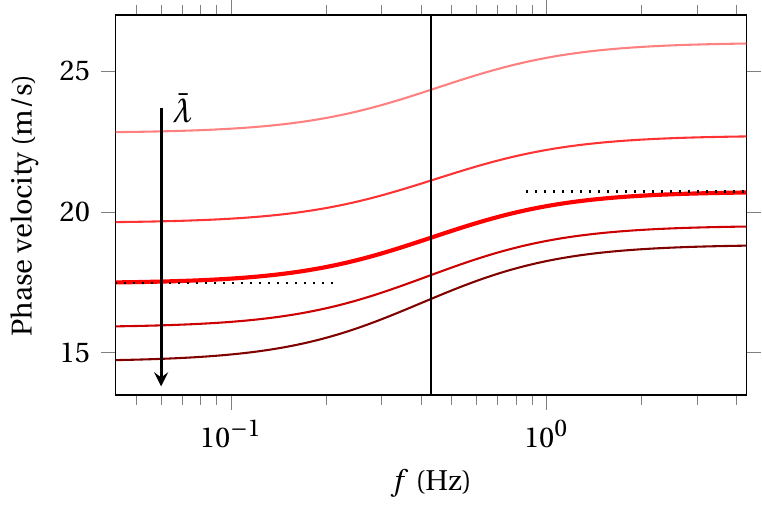}
	\end{minipage}
	\hfill
	\begin{minipage}{0.49\textwidth}
		\centering
		(b)
		
		\includegraphics{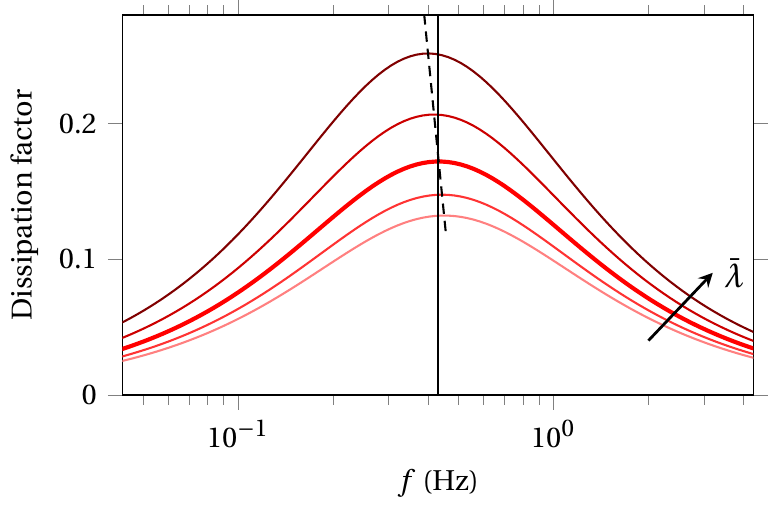}
	\end{minipage}
	
	\caption{Dispersive plane waves propagating in the transverse direction to tensile stretch. Evolution of the phase velocity (a) and of the dissipation factor (b) in terms of the (logarithmic) frequency $f = \omega/(2\pi)$ for the longitudinal stretches $\bar\lambda$ in $\{0.6,0.8,\dots ,1.4\}$. Thick red lines mark the undeformed case $\bar\lambda = 1$. The dashed line in (b) marks the locus of maximum dissipation. \label{fig:Dispersive}}
\end{figure}

Dispersion and dissipation properties can be deduced from the dispersion relationship $\rho {\omega^2}/{\kappa^2} = \mu_x$, see \citet{carcione15} for complements. Using the above expression of the dynamic modulus $\mu_x$, one deduces the \emph{phase velocity}
\begin{equation}
	\frac{\omega}{\text{Re}\,\kappa} = \pm\sqrt{ \frac{2\, (1+D^2)}{1+\sqrt{1+D^2}} } \sqrt{ \frac{|\text{Re}\, \mu_x|}{\rho} }
\end{equation}
and \emph{dissipation factor}
\begin{equation}
	D = -\frac{\text{Im}(\kappa^2)}{\text{Re}(\kappa^2)} = \frac{\text{Im}\, \mu_x}{\text{Re}\, \mu_x} = D_{0} \frac{ 2\Omega\Omega_0  }{\Omega^2 + \Omega_0^2} 
	\label{DissipFact}
\end{equation}
where $\Omega = \omega\tau$ is a normalised frequency, and
\begin{equation}
	D_{0} = \frac{g}{2\Omega_0} \frac{\bar\mu_x^\text{v}}{\bar\mu_x^\text{v} + (1-g) [\bar{T}^\text{e}_\text{d}]_{11} } \, ,\qquad
	\Omega_0^2 = (1-g)\frac{ \bar\mu_x^\text{v} + [\bar{T}^\text{e}_\text{d}]_{11} }{ \bar\mu_x^\text{v} + (1-g) [\bar{T}^\text{e}_\text{d}]_{11} }
\end{equation}
for any applied stretch $\bar \lambda$.
The frequency evolution of the medium's phase velocity and dissipation factor for several levels of pre-deformation is displayed in Figure~\ref{fig:Dispersive}. A first look at these curves shows that the phase velocity is frequency-dependent (Fig.~\ref{fig:Dispersive}a), i.e. the material is dispersive, and that the dissipation factor has a bell-shaped curve in terms of logarithmic frequencies (Fig.~\ref{fig:Dispersive}b). According to Eq.~\eqref{DissipFact}, the dissipation factor reaches its maximum value $D_{0}$ at the normalised frequency $\Omega=\Omega_0$.

The thick line in Fig.~\ref{fig:Dispersive} marks the undeformed state, where the tensile equilibrium stretch $\bar \lambda$ equals unity. In this case, the high-frequency limit of the phase velocity is the shear wave speed $c$, and its low-frequency limit equals $\Omega_0 c$ with $\Omega_0 = \sqrt{1-g}$. These asymptotes are marked by horizontal dotted lines in Fig.~\ref{fig:Dispersive}a. The dissipation factor reaches its maximum value $D_0 = g/(2\Omega_0) \approx 0.17$ at the normalised frequency $\Omega_0$. Given the numerical values deduced from Table~\ref{tab:Params}, the undeformed material is mostly attenuating about the frequency $\Omega_0/(2\pi\tau) = 0.43$~Hz (vertical line in Fig.~\ref{fig:Dispersive}), i.e. in the low-frequency range.

Now, let us consider several levels of applied pre-deformation by varying $\bar\lambda$. Figure \ref{fig:Dispersive}a shows that the phase velocity decreases monotonically with stretch. In the vicinity of the undeformed state, Eq.~\eqref{DissipFact} tells us that the maximum dissipation $D_0$ increases with the pre-stretch $\bar\lambda$, whereas $\Omega_0$ decreases with increasing pre-stretch. This evolution is confirmed in Fig.~\ref{fig:Dispersive}b, where the bell-shaped curves flatten with decreasing stretch. Moreover, these curves are slightly shifted towards decreasing frequency when the stretch is increased, as shows the dashed curve marking the locus of maximum dissipation. In conclusion, the elongated material's dissipation occurs at lower frequency and with more significant effect than in the compressed material.\footnote{The opposite tendencies are observed if the incremental waves propagate in the direction of elongation (not shown here). Indeed, if the material is elongated in a given direction, then it is simultaneously compressed in the transverse directions due to the incompressibility property.}

\section{Application to a pre-stressed phononic crystal}\label{sec:Phononic}

In this section, we consider the periodic structure proposed in \citet{barnwell16} in view of adapting their study to the viscoelastic case. Thus, we follow a similar approach based on Bloch wave analysis (aka. plane-wave expansion method) to study the system's effective dynamic response. In particular, the computation of the static pre-deformation described hereinafter is not original. Nevertheless, the theory introduced in previous sections suggests that the consideration of viscoelastic dissipation modifies substantially the analysis of incremental motions.

\begin{figure}
	\centering
	\includegraphics{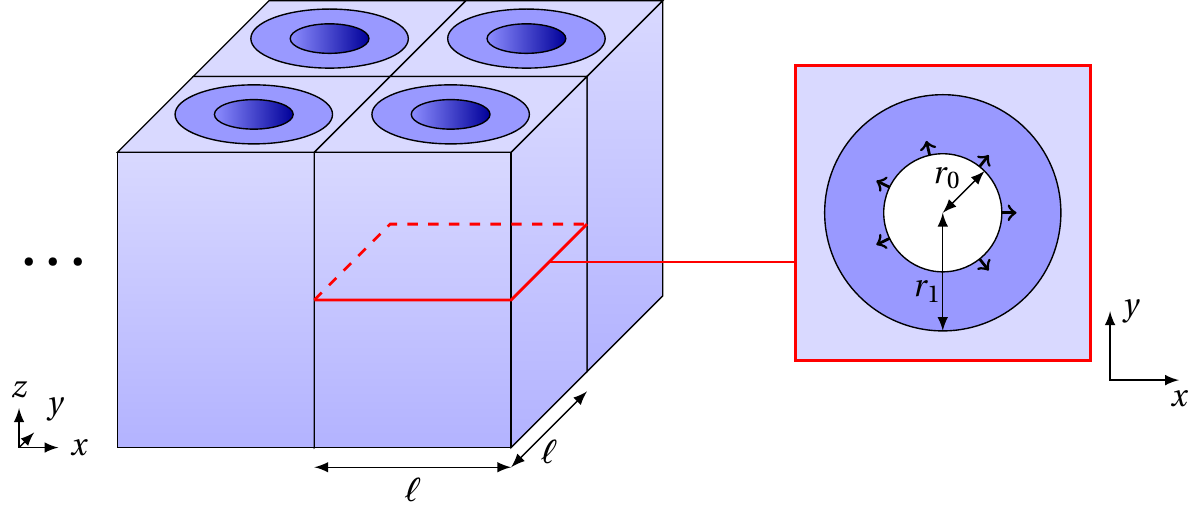}
	
	\caption{Infinite periodic structure from \citet{barnwell16}, and its square unit cell. Only the inner radius of the annulus region varies with pre-deformation, due to vertical stretching of the hollow cylinders combined with application of a well-chosen inner pressure. \label{fig:Barnwell}}
\end{figure}

As represented in Fig.~\ref{fig:Barnwell}, the phononic crystal at hand consists of a two-dimensional periodic structure with square unit cells (the structure is assumed invariant along the $z$-axis). Each unit cell has an embedded annulus region made of a soft rubber-like material corresponding to hollow cylinders in three-dimensional space. The inner region of the annuli is filled with an inviscid gas allowing to control the pressure inside the cylinders, while the outer region consists of another solid material (host material). All unit cells are submitted to the same static pre-deformation, resulting from an applied inner pressure combined with cylinder elongation along $z$. This elongation is chosen in such a way that the outer region is initially undeformed{\,---\,}in other words, pre-deformation is restricted to the annulus region.

In the present study, the cylinders are assumed to be made of a viscoelastic Mooney--Rivlin material with $n=1$ relaxation mechanism, which reference parameters are given in Table~\ref{tab:Params}. Moreover, we assume that the host material has the same relaxation function as the cylinders, i.e. the same parameters $g$, $\tau$. In what follows, elastic behaviour is thus recovered as a special case, namely $g=0$ or $\tau \to +\infty$. Let us first derive the equations governing incremental wave propagation within a single unit cell before the full periodic structure is addressed.

\subsection{Pre-deformed unit cell}

Let us consider the cylindrical coordinate system of a cell (see Fig.~\ref{fig:Barnwell}), which deforms exclusively in the annulus region $r_0 \leq \bar r\leq r_1$. In a standard fashion, we introduce the coordinates $(\bar r,\bar \theta,\bar z)$ and $(R,\Theta,Z)$ of a particle in the deformed and undeformed states, respectively (see picture in Fig.~\ref{fig:Config}). The deformation includes a radial component $\bar r\mapsto R(\bar r)$ and a vertical component $Z = \bar z/\zeta$ where $\zeta$ is constant. The corresponding deformation gradient tensor reads $\bar{\bm F} = \text{diag}\left[\bar \lambda_i\right]$ for $i\in \lbrace r, \theta, z\rbrace$, with the radial stretch $\bar\lambda_r = 1/R'(\bar r)$,  angular stretch $\bar\lambda_\theta = \bar r/R(\bar r)$, and vertical stretch $\bar\lambda_z = \zeta$. The incompressibility constraint \eqref{Incomp} for $\bar{\bm F}$ implies 
\begin{equation}\label{radii_law}
	R(\bar r)^2 = \zeta \bar r^2 + (1-\zeta) r_1^2,
	\qquad
	\frac{r_1^2 - R(r_0)^2}{r_1^2} < \zeta < \frac{r_1^2}{r_1^2 - r_0^2} ,
\end{equation}
where we have used the fact that the outer annulus boundary $r_1 = R(r_1)$ is invariant. In Eq.~\eqref{radii_law}, the bounds for the vertical stretch $\zeta$ follow from the requirement of (real) achievable positions for the inner boundary $R(r_0) > 0$ with $r_0 > 0$. Thus, imposing the position $R(r_0)$ of the inner boundary yields the value of the vertical stretch, and vice versa. The expression of the other stretches $\bar\lambda_r$, $\bar\lambda_\theta$ in terms of $\bar r$ then follows from this boundary condition.

The equilibrium equation \eqref{StaticStress} involves the static stress $\bar{\bm T}$ with the elastic response $\bar{\bm T}^\text{e}$ of Eq.~\eqref{Elast}. Here, it is more convenient to rewrite the elastic stress as $\bar{\bm T}^\text{e} = (\partial \bar W/\partial \bar{\bm F})\, \bar{\bm F}^\transpose\!$, where the strain energy $\bar W$ is expressed in terms of the $\bar r$-dependent stretches $\bar\lambda_i$ defined above. Thus, the elastic response $\bar{\bm T}^\text{e} = \text{diag}\big[\bar\lambda_i\, \partial\bar W/\partial \bar\lambda_i\big]$ is a diagonal tensor, where the partial derivatives of $\bar W$ with respect to the stretches are radial functions. Finally, up to a redefinition of pressure, the equilibrium equations in cylindrical coordinates tell us that the Lagrange multiplier depends on $\bar r$ only, and so does the radial stress component $\bar{T}_{rr}$ as well. The latter is given by
\begin{equation}
	\begin{aligned}
		\bar{T}_{rr}(\bar r) &= -p_0 + (1 - g) \int_{r_0}^{\bar r} \frac{1}{\varrho} \left(\bar\lambda_\theta(\varrho) \frac{\partial\bar W}{\partial \bar\lambda_\theta}(\varrho) - \bar\lambda_r(\varrho) \frac{\partial\bar W}{\partial \bar\lambda_r}(\varrho)\right) \text{d}\varrho \\
		&= -p_0 + (1-g) \left(\frac{\mathfrak{C}_1}{\zeta} + \zeta\mathfrak{C}_2\right) \left[\ln\left(\frac{R(\bar r)^2/\bar r^2}{R(r_0)^2/r_0^2}\right) + \frac{1 - \zeta }{\zeta}\left(\frac{r_1^2}{\bar r^2} - \frac{r_1^2}{r_0^2}\right)\right] ,
	\end{aligned}
	\label{BarnwellRadial}
\end{equation}
where the pressure $\bar{T}_{rr}(r_0) = -p_0$ is imposed at the inner surface of the annulus region. For a given input pressure $p_0$, the vertical stretch $\zeta$ is chosen in such a way that the integral term of Eq.~\eqref{BarnwellRadial} equals $p_0$ at the radius $\bar r = r_1$. Therefore, $\zeta$ is solution to a transcendental equation which can be solved using Newton's method. This way, the radial traction $\bar{T}_{rr}$ vanishes at the outer boundary of the annulus region, which remains invariant under such a static pre-deformation.

With this pre-deformation, we now consider time-harmonic incremental displacements $\ubar{\bm u} = \hat{\bm u}\, \text{e}^{\text{i}\omega t}$ of the form $\hat{\bm u} = w(r,\theta)\, \bm{e}_z$, for which the incompressibility constraint is always satisfied. In cylindrical coordinates, the time-harmonic displacement gradient tensor $\hat{\bm H}$ has components $w_{,r}$ and $w_{,\theta}/r$ along $\bm{e}_z \otimes \bm{e}_r$ and $\bm{e}_z \otimes \bm{e}_\theta$, respectively. Incremental waves are governed by Eq.~\eqref{IncrementMot}, i.e. $-\rho\omega^2 \hat{\bm u} = \text{div}\, \hat{\bm \Sigma}$, where the time-harmonic incremental stress $\hat{\bm \Sigma}$ is deduced from Eq.~\eqref{IncrementStressDyn}, cf. previous paragraph for the description of the equilibrium state (quantities with overbar). In cylindrical coordinates, the radial and angular components make the Lagrange multiplier vanish. The vertical component yields the harmonic wave equation
\begin{equation}
	-\rho\omega^2 w = \frac{1}{r}\frac{\partial}{\partial r}\left(r \hat{\Sigma}_{zr} \right) + \frac{1}{r}\frac{\partial}{\partial \theta} \hat{\Sigma}_{z\theta}\, ,
	\qquad \hat{\Sigma}_{zr} = \mu_r \frac{\partial w}{\partial r} \, ,
	\qquad \hat{\Sigma}_{z\theta} = \frac{\mu_\theta}{r}\frac{\partial w}{\partial \theta}
	\label{IncrementBarnwell}
\end{equation}
with the coefficients
\begin{equation}
	\mu_p = (1-g) [\bar T^\text{e}_\text{d}]_{pp} + \left(1 - \frac{g}{1+\text{i}\omega\tau}\right) \bar\mu_p^\text{v} \, ,\qquad
	p = r,\theta
	\label{CoeffsBarnwell}
\end{equation}
where
\begin{equation}
	[\bar T^\text{e}_\text{d}]_{pp} = \frac23 \left[\left(3\bar \lambda_p^2 - \bar I_1\right) \mathfrak{C}_1 + \left(3\bar I_1 \bar \lambda_p^2 - 3 \bar \lambda_p^4 - 2 \bar I_2\right) \mathfrak{C}_2\right]
\end{equation}
and
\begin{equation}
	\bar\mu_r^\text{v} = \frac23 \left[\bar I_1 \mathfrak{C}_1 + (2 \bar I_2 - 3 \bar\lambda_\theta^{-2}) \mathfrak{C}_2\right] ,\qquad
	\bar\mu_\theta^\text{v} = \frac23 \left[\bar I_1 \mathfrak{C}_1 + (2\bar I_2 - 3 \bar\lambda_r^{-2}) \mathfrak{C}_2\right] .
\end{equation}
One observes that these expressions have a very similar form to that in Sec.~\ref{subsec:Illustr}, which follows from the fact that the static deformation gradient is diagonal. In the relaxed elastic limit $\omega\to 0$, the above expressions match Eq.~(7) of \citet{barnwell16}, where the above coefficients reduce to $\mu_p = 2 (1-g) \big(\mathfrak{C}_1\bar\lambda_p^2 + \mathfrak{C}_2\bar\lambda_z^{-2}\big)$ for $p = r, \theta$. Note however that the unrelaxed elastic limit $\omega \to +\infty$ yields different expressions. This observation is consistent with the fact that a high-frequency incremental perturbation couples both relaxed and unrelaxed elastic solid limits (in the static pre-deformation and its dynamic perturbation, respectively).

\subsection{Pre-deformed periodic structure}

We rely on the plane-wave-expansion method to analyse the band gap structure of the pre-deformed periodic material described in Fig.~\ref{fig:Barnwell}. We consider time-harmonic antiplane waves polarised along $z$, corresponding to incremental displacements $\ubar{\bm u} = \hat{\bm u}\, \text{e}^{\text{i}\omega t}$ of the form $\hat{\bm u} = w(x,y)\, \bm{e}_z$ where $\omega$ is the angular frequency. The motion is governed by the incremental wave equation $-\rho\omega^2 \hat{\bm u} = \text{div}\, \hat{\bm \Sigma}$ with the incremental stress tensor of Eq.~\eqref{IncrementStressDyn}, where pre-deformation is described by the quantities with overbars.

Due to the periodicity of the system, the static pre-deformation is periodic with period $\ell$ in both directions $x$, $y$. Following Bloch's theorem, we seek wave fields of the form
\begin{equation}
	w({\bm x}) = \text{e}^{\text{i}\, {\bm \kappa}\cdot {\bm x}} \left(\sum_{\bm G} \mathcal{W}[{\bm G}]\, \text{e}^{\text{i}\, {\bm G} \cdot {\bm x} }\right) ,
	\label{Bloch}
\end{equation}
where the Bloch wavevector $\bm \kappa$ and reciprocal lattice vectors $\bm G$ are orthogonal to the vertical $z$-axis. Reciprocal lattice vectors span the integer combinations of the primitive vectors $\bm{b}_j = \frac{2\pi}{\ell} \bm{e}_j$, i.e. we may write $\bm{G} = m_j \bm{b}_j$ with integer components $(m_1, m_2)$ in $\mathbb{Z}^2$. Therefore, inside the parentheses, a $\ell$-periodic function in two dimensions is represented by its Fourier series with coefficients $\mathcal{W}[{\bm G}]$. In fact, translation of $\bm{x}$ by any lattice vector $\bm{R}$ defined as an integer combination of the primitive vectors $\bm{a}_i = \ell \bm{e}_i$ keeps the bracketed function invariant. We note that the relationship $\bm{a}_i \cdot \bm{b}_j = 2\pi\, \delta_{ij}$ between primitive vectors is satisfied.

We now transform the incremental wave equation \eqref{IncrementBarnwell} governing vertical displacements to Cartesian coordinates using the change of variables $(x, y) = r (\cos\theta, \sin\theta)$. Over each unit cell, Eq.~\eqref{IncrementBarnwell} rewrites as
\begin{equation}
	-\rho\omega^2 w = \frac{\partial}{\partial x} \hat{\Sigma}_{zx} + \frac{\partial}{\partial y} \hat{\Sigma}_{zy} \, ,
	\label{IncrementCartesian}
\end{equation}
where
\begin{equation}
	\begin{bmatrix}
		\hat{\Sigma}_{zx}\\
		\hat{\Sigma}_{zy}
	\end{bmatrix} =
	\frac1{x^2+y^2}
	\begin{bmatrix}
		x^2 \mu_r + y^2 \mu_\theta & xy (\mu_r - \mu_\theta)\\
		xy (\mu_r - \mu_\theta) & y^2 \mu_r + x^2 \mu_\theta
	\end{bmatrix}
	\begin{bmatrix}
		w_{,x}\\
		w_{,y}
	\end{bmatrix} = {\bm A}\, {\bm h}
	\label{CoeffXYBarnwell}
\end{equation}
and the vector ${\bm h}$ gathers the displacement gradient components $w_{,x}$, $w_{,y}$.

Next, the expressions of $\rho$ and ${\bm A}$ are extended to the whole periodic structure by spatial periodisation, i.e. the coefficients of the wave equation \eqref{IncrementCartesian} are rewritten as Fourier series
\begin{equation}
	\rho({\bm x}) = \sum_{\bm G} \mathcal{R}[\bm{G}]\, \text{e}^{\text{i}\, {\bm G} \cdot {\bm x} }, \qquad
	{\bm A}({\bm x}) = \sum_{\bm G} \bm{\mathcal{A}}[\bm{G}]\, \text{e}^{\text{i}\, {\bm G} \cdot {\bm x} } ,
	\label{FourierSeries}
\end{equation}
which coefficients are given by
\begin{equation}
	\mathcal{R}[\bm{G}] = \frac1{\ell^2} \int_{\text{Cell}}
	\rho({\bm x})\, \text{e}^{-\text{i}\, \bm{G}\cdot \bm{x}} \text{d}{\bm x}
	\, , \qquad
	\bm{\mathcal{A}}[\bm{G}] = \frac1{\ell^2} \int_{\text{Cell}}
	{\bm A}({\bm x})\, \text{e}^{-\text{i}\, \bm{G}\cdot \bm{x}} \text{d}{\bm x} \, .
	\label{FourierCoeffs}
\end{equation}
Similarly, the vector $\bm h$ has the Fourier coefficients $\bm{\mathcal{H}}[\bm{G}] = \text{i}\, \mathcal{W}[\bm{G}] \left(\bm{G}+\bm{\kappa}\right)$ deduced from the Bloch wave Ansatz \eqref{Bloch}. Finally, substitution of the Fourier series representations \eqref{FourierSeries} in Eq.~\eqref{IncrementCartesian} yields the algebraic problem
\begin{equation}
	\left({\bf K} - \omega^2 {\bf M}\right) {\bf w} = {\bf 0}
	\label{BlochEigen}
\end{equation}
\begin{equation*}
	\text{with}\qquad
	{\text K}_{{\bm G}' {\bm G}} = (\bm{G}+\bm{\kappa})^\transpose  \bm{\mathcal{A}}[\bm{G}'-\bm{G}] \, (\bm{G}'+\bm{\kappa}) ,\qquad
	{\text M}_{{\bm G}' {\bm G}} = \mathcal{R}[\bm{G}'-\bm{G}] ,
\end{equation*}
where the vector ${\bf w} = (\mathcal{W}[\bm{G}])^\transpose$ gathers the lattice Fourier coefficients of the vertical displacement $w$. Here, we have used the convolution theorem of Fourier series, and the product rule $\bm{\nabla}\cdot(\bm{A}\bm{h}) = \bm{h}^\transpose \bm{\nabla}\cdot \bm{A}  + \bm{A}:\bm{\nabla h}$ with ${\bm A} = {\bm A}^\transpose$ for the Cartesian divergence operator in two space dimensions.

Up to the notations used in the present study, Eq.~\eqref{BlochEigen} matches exactly Eq.~(23) of the study by \citet{barnwell16}. However, for a fixed wavevector ${\bm\kappa}$, Eq.~\eqref{BlochEigen} can no longer be viewed as a generalised eigenvalue problem for $\omega^2$. In fact, according to the expression of the coefficients $\mu_r$, $\mu_\theta$ in Eq.~\eqref{CoeffsBarnwell}, the matrix $\bm{A}$ with lattice Fourier coefficients $\bm{\mathcal{A}}[\bm{G}]$ is now complex-valued and dependent on the angular frequency $\omega${\,---\,}and so does the matrix $\bf K$ as well. Nevertheless, non-trivial solutions to Eq.~\eqref{BlochEigen} still express the singularity of the matrix ${\bf K} - \omega^2 {\bf M}$, e.g. to be evaluated with respect to $\omega = \omega(\bm{\kappa})$ at some fixed wavevector $\bm\kappa$.\footnote{Alternatively, one might seek values of ${\bm\kappa} = {\bm\kappa}(\omega)$ such that the matrix ${\bf K} - \omega^2 {\bf M}$ becomes singular for some fixed value of $\omega$.  Both approaches are usually found to be equivalent (see \citet{andreassen2013, li21} for instance).} For any wavevector $\bm{\kappa} = \kappa\bm{n}$ with complex wavenumber $\kappa$, the absolute phase velocity of Bloch waves deduced from Eq.~\eqref{BlochEigen} equals $\text{Re}\,\omega/\text{Re}\,\kappa$, while the attenuation in space and time is given by $-\text{Im}\, \kappa$ and $\text{Im}\, \omega$, respectively.

In the elastic case ($g=0$), the evolution of the dispersion curves with applied pre-deformation and selected material parameters is well-described in literature \citep{depascalis2020}. In the viscoelastic case ($g\neq0$), performing the Bloch wave analysis is more involved due to complex values and frequency-dependence in Eq.~\eqref{BlochEigen}. To address these challenges, several approaches are adopted in literature, including direct computational methods \citep{wang15,krushynska16} as well as dedicated algorithms \citep{zhao09,mokhtari19}. In the next subsection, we introduce a perturbation method based on the small parameter $g$, that does not involve very sophisticated algorithms{\,---\,}see technical details in Section~\ref{subsec:Results}. We then investigate separately the influence of $g$ and $\tau$ on wave dispersion at some applied stretch $\zeta\geq1$.

\subsection{Perturbation theory}\label{subsec:Perturb}

When solving the algebraic problem \eqref{BlochEigen} with respect to $\omega$ by means of a given numerical method, it is often useful to provide an initial guess for $\omega$. To do so, let us introduce a perturbation method based on the small parameter $g$ to approximate the angular frequency. For this purpose, we introduce generalised eigenvectors ${\bf r}$ forming a basis of the right null space of ${\bf K} - \omega^2 {\bf M}$.

We seek solutions to Eq.~\eqref{BlochEigen} in the form of power series of $g$, i.e. we set $\omega \simeq \omega^0 + g \omega^1$ and ${\bf r} \simeq {\bf r}^0 + g {\bf r}^1$, where the zeroth-order quantities $\omega^0$, ${\bf r}^0$ correspond to $g=0$. Similar expansions for $\bf K$ and $\bf M$ are thus introduced, where the Hermitian matrix $\bf M$ is found to be independent on $g$. As shown in the expression \eqref{CoeffsBarnwell}-\eqref{CoeffXYBarnwell} of $\bm A$, the matrices $\bm{\mathcal{A}}$ and ${\bf K}$ can be linearised with respect to $g$, leading to perturbations of the form $\bm{\mathcal{A}} \simeq \bm{\mathcal{A}}^0 + g \bm{\mathcal{A}}^1$ and ${\bf K} \simeq {\bf K}^{0} + g {\bf K}^{1}$, where the zeroth-order matrices corresponding to the elastic limit $g=0$ are Hermitian. The first-order matrix ${\bf K}^{1}$ has the coefficients
\begin{equation}
	{\text K}_{{\bm G}' {\bm G}}^1 = \bm{\kappa}^{1\transpose} \bm{\mathcal{A}}^0[\bm{G}'-\bm{G}] \, (\bm{G}'+\bm{\kappa}^0) + (\bm{G}+\bm{\kappa}^0)^\transpose  \bm{\mathcal{A}}^1[\bm{G}'-\bm{G}] \, (\bm{G}'+\bm{\kappa}^0) + (\bm{G}+\bm{\kappa}^0)^\transpose  \bm{\mathcal{A}}^0[\bm{G}'-\bm{G}] \, \bm{\kappa}^1
	\label{PertMatrix}
\end{equation}
following from a power-series expansion of the wavevector $\bm{\kappa} \simeq \bm{\kappa}^0 + g \bm{\kappa}^1$, where the first-order matrix $\bm{\mathcal{A}}^1$ is dependent on $\omega^0$.
Injecting this Ansatz in the algebraic problem \eqref{BlochEigen} leads to the conditions
\begin{equation}
	\begin{aligned}
		\text{order 0:} \qquad &\left({\bf K}^0 - (\omega^0)^2 {\bf M}\right){\bf r}^0 = {\bf 0} \, ,\\
		\text{order 1:} \qquad &\left({\bf K}^1 - 2\omega^0\omega^1 {\bf M}\right){\bf r}^0 + \left({\bf K}^0 - (\omega^0)^2 {\bf M}\right){\bf r}^1 = {\bf 0} \, ,
	\end{aligned}
	\label{PerturbationCond}
\end{equation}
at zeroth order and first order of the small parameter $g$.

Now, we left-multiply the second line of Eq.~\eqref{PerturbationCond} by the vector ${\bf r}^{0\dagger}$ where the dagger symbol denotes the transpose conjugate. At the same time, we compute the transpose conjugate of the first line of Eq.~\eqref{PerturbationCond}, recalling that ${\bf K}^0$ and $\bf M$ are Hermitian matrices. Thus, combining both identities leads to the following approximate expression of the angular frequency
\begin{equation}
	\omega \simeq \omega^0 + g\omega^1 = \omega^0 + \frac{g}{2 \omega^0}  \frac{{\bf r}^{0\dagger} {\bf K}^1 {\bf r}^0}{ {\bf r}^{0\dagger} {\bf M}\, {\bf r}^0}
	\label{PerturbApprox}
\end{equation}
at first order in $g$. One observes that the increment of the angular frequency is linear with respect to the (presumably small) perturbation $g {\bf K}^{1}$ of the matrix ${\bf K}$. By construction, the truncation error $| \omega^0 + g \omega^1 - \omega|$ introduced by the first-order perturbation is necessarily of order $O(g^2)$. Upon division by $\omega$, the same rate of convergence is obtained for the relative error $| (\omega^0 + g \omega^1)/\omega - 1|$.

\paragraph{Illustration}

Let us go back to the example studied in Section~\ref{subsec:Illustr} where the dispersion relationship takes the form $K = \rho\omega^2$ with $K = \mu_x\kappa^2$. Similarly, we are considering perturbed quantities in terms of the small parameter $g$, where the elastic case corresponds to $g = 0$. Thus, we seek $\omega$ in the form of a power series in $g$, where we have assumed $\kappa \simeq \kappa^0 + g \kappa^1$ and $\mu_x \simeq \mu_x^0 + g \mu_x^1$. According to Eq.~\eqref{Dispersion}, the zeroth-order term of the dynamic modulus $\mu_x$ equals $\mu_x^0 = [\bar{T}^\text{e}_\text{d}]_{11} + \bar\mu_x^\text{v}$, and the first-order term $\mu_x^1$ is an $\omega^0$-dependent complex number. We find the relationship $K^0 = \rho(\omega^0)^2$ at order zero, and
\begin{equation}
	\omega \simeq \omega^0 + \frac{g K^1}{ 2\omega^0\rho }
	\qquad\text{with}\qquad
	K^1 = 2 \mu_x^0 \kappa^0 \kappa^1 + (\kappa^0)^2 \mu_x^1
	\label{PerturbExample}
\end{equation}
at order one. Note the similarity with the case of the phononic crystal \eqref{PerturbApprox}.

\begin{figure}
	\begin{minipage}{0.49\textwidth}
		\centering
		(a)
		
		\includegraphics{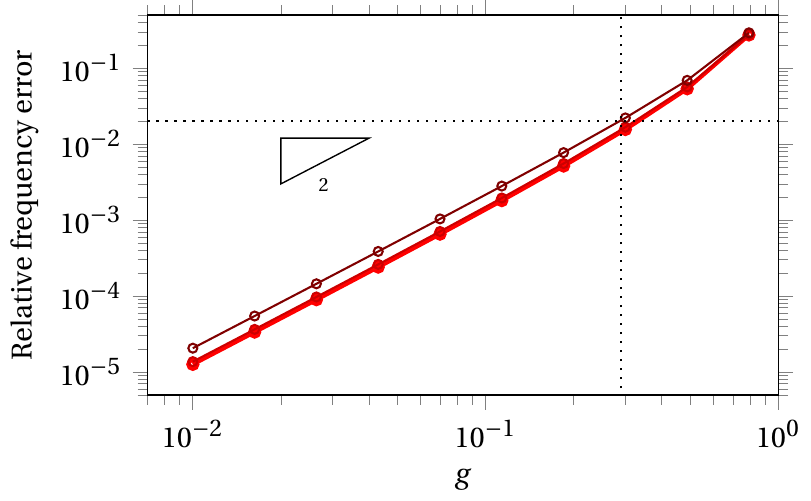}
	\end{minipage}
	\hfill
	\begin{minipage}{0.49\textwidth}
		\centering
		(b)
		
		\includegraphics{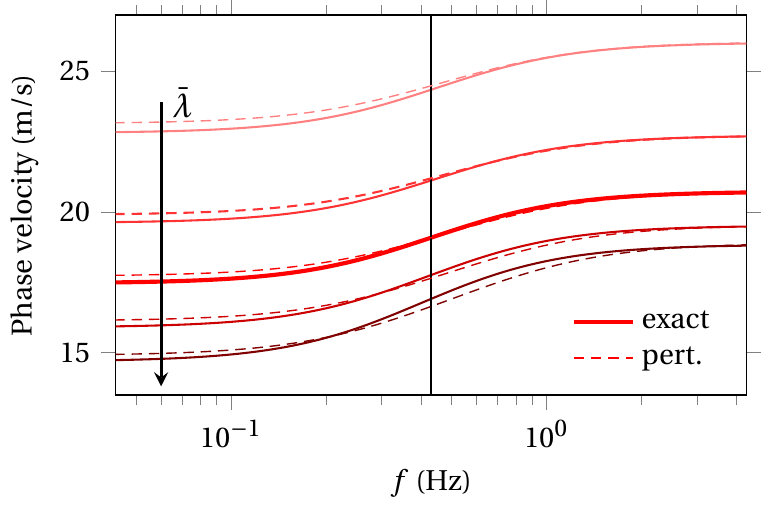}
	\end{minipage}
	
	\caption{Validation of the perturbation theory \eqref{PerturbExample} where $g$ is a small parameter. For a given complex wavenumber related to Fig.~\ref{fig:Dispersive}, the frequency computed using the perturbation method is compared to its exact value deduced from dispersion analysis. (a) Evolution of the relative frequency error at the frequency $f=\Omega_0/(2\pi \tau)$ of maximum dissipation, where each curve corresponds to a value of stretch $\bar\lambda$ in $\{0.6,0.8,\dots, 1.4\}$. (b) Dispersion errors for $g = 0.29$: exact curve against first-order perturbation. \label{fig:PerturbError}}
\end{figure}

To evaluate the error introduced by the present first-order approximation, we consider an exact wavenumber $\kappa$ that was obtained by solving the dispersion relationship $\rho \omega^2/\kappa^2 = \mu_x$ for a given real frequency $\omega$. By setting $\kappa^0 = \text{Re}\, \kappa$ and $g\kappa^1 = \text{i}\, \text{Im}\, \kappa$, the perturbation \eqref{PerturbExample} produces an approximation of the angular frequency $\omega$, introducing a relative error of order $O(g^2)$.
This property is illustrated in Figure~\ref{fig:PerturbError}a (black triangle), where an appropriate rate of decay in log-log coordinates is found for all the applied stretches $\bar \lambda$ at the frequency of maximum dissipation. Note that the relative error does not exceed $\approx 2$\% with the value $g = 0.29$ of Table~\ref{tab:Params} (vertical dotted line) at the frequency of maximum dissipation.

Fig.~\ref{fig:PerturbError}b displays the dispersion error introduced by the perturbation method for the value $g = 0.29$ of Table~\ref{tab:Params}. Here, we compare the dispersion curves of Fig.~\ref{fig:Dispersive}a with the same quantity obtained by perturbation \eqref{PerturbExample}. One observes that the overall evolution is well-reproduced, and that the perturbation method performs best in the high-frequency range. In the present configuration, phase velocities are slightly overestimated in compression and underestimated in elongation at the frequency of maximum dissipation (about the vertical line in Fig.~\ref{fig:Dispersive}b). Computation of the corresponding relative frequency errors shows that they do not exceed 2\% over the frequency range of the figure.

\subsection{Results and discussion}\label{subsec:Results}

We present the results obtained for the phononic crystal in dimensionless form by setting
\begin{equation}
	\check{\bm{\kappa}} = \ell \bm{\kappa} , \qquad \check{\omega} = \frac{\ell \omega}{c_0}, \qquad
	\check{\bm{x}} = \frac{\bm{x}}{\ell}, \qquad \check\rho(\check{\bm{x}}) = \frac{\rho(\bm{x})}{\rho_0}, \qquad
	\check{\mathfrak{C}}_k(\check{\bm{x}}) = \frac{\mathfrak{C}_k(\bm{x})}{\mu_0} \quad (k=1,2), \qquad \check{\tau}=\frac{c_0 \tau}{\ell},
	\label{Scaling}
\end{equation}
where $\ell$ is the cell size, and $\rho_0$, $\mu_0 = 2(\mathfrak{C}_{10} + \mathfrak{C}_{20})$, $c_0 = \sqrt{\mu_0/\rho_0}$ are mechanical properties of the host material deduced from Table~\ref{tab:Params} ($\bar r>r_1$). In the following, $\check R_0=0.3$ and cylindrical inclusions are thought softer than the host so that the imposed pre-deformation is achievable \citep{barnwell16,depascalis2020}. To this end we simply assume a uniform softening by setting $\mathfrak{C}_{11}/\mathfrak{C}_{10} = \mathfrak{C}_{21}/\mathfrak{C}_{20} = \rho_1/\rho_0 = 0.1$ for the cylinder material ($r_0<\bar r<r_1$), while  the relaxation function is the same in both regions (parameters $g$, $\tau$ of Table~\ref{tab:Params}). We neglect the added-mass effect caused by the presence of air in the cylinder core by setting the mechanical parameters to zero in the corresponding region ($\bar r<r_0$).

\begin{figure}
	\centering
	\includegraphics[scale=0.4]{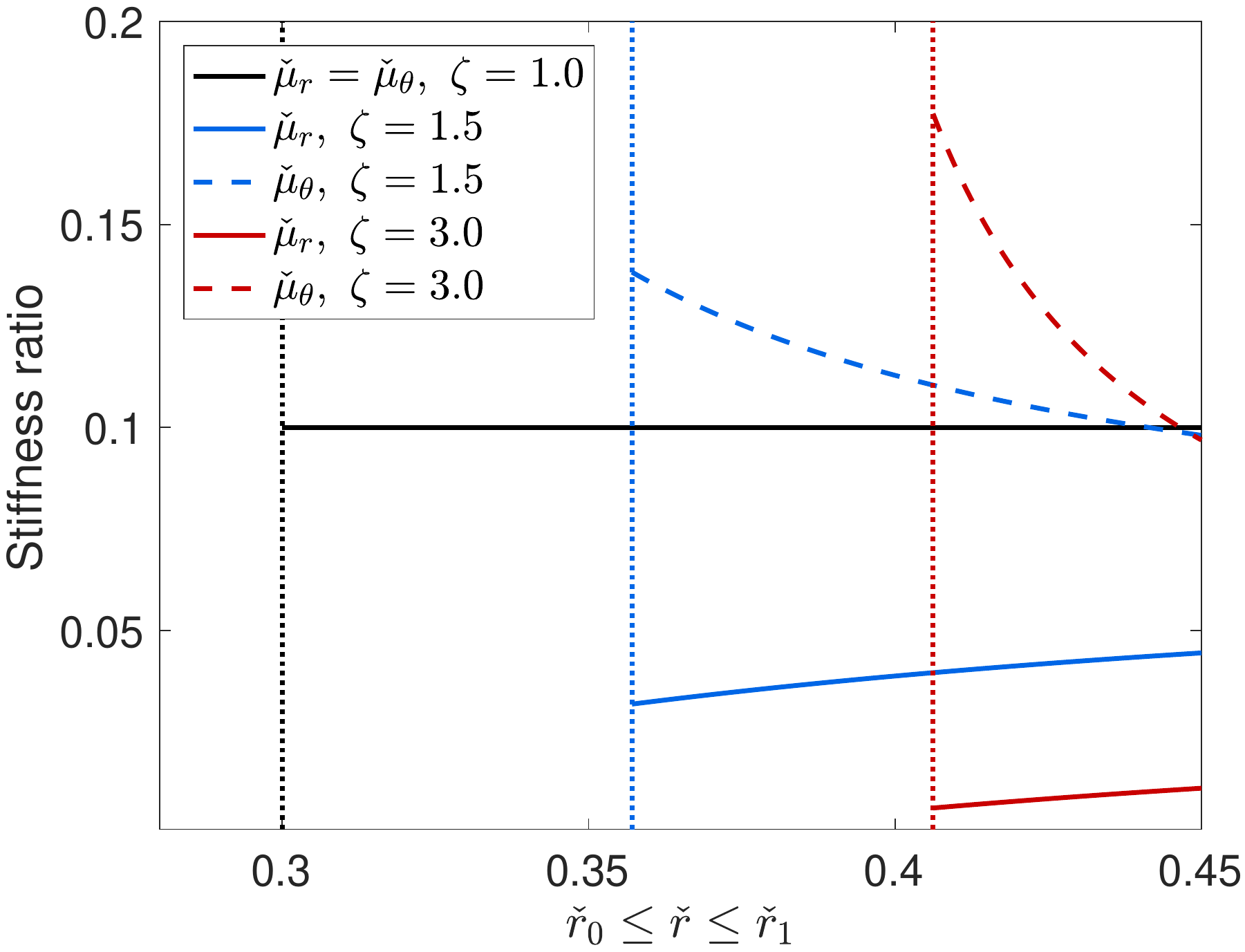}
	
	\caption{Static deformation at imposed stretch $\zeta$. Evolution of the radial and angular nondimensional incremental stiffnesses at equilibrium in pre-stressed Mooney--Rivlin cylinders with softness ratio $0.1$.}
	\label{fig:MR_prestress}
\end{figure}

To avoid any instability that can occur to compressed hollow cylinders \citep{goriely_vandiver_destrade2008,dpr11} and which might lead to consequent dramatic change of the periodic structure, the pre-deformation applied to the cylindrical annuli is of extensional type, i.e. $\zeta>1$.  Fig.~\ref{fig:MR_prestress} illustrates the effect of the pre-deformation applied to the periodic material by showing the radial evolution of the incremental shear moduli in the cylinders at various stretches $\zeta$ in the elastic case ($g=0$). Vertical dotted lines mark the inner radius $\check r_0$ of the cylinders ($\check r_0=0.3$, $\check{p}_0\equiv p_0/\mu_0= 0$ at $\zeta=1$; $\check r_0\approx0.36$, $\check{p}_0 \approx0.19$ at $\zeta=1.5$; and $\check r_0\approx0.41$, $\check{p}_0 \approx0.16$ at $\zeta=3$ according to \eqref{radii_law} whilst outer radius  is fixed at  $\check r_1=0.45$).

To investigate the effects of viscoelastic dissipation on the dispersion properties, we apply the perturbation method described in Sec.~\ref{subsec:Perturb}, providing a $\check{\omega}(\check{\bm{\kappa}})$-method for the approximate resolution of Eq.~\eqref{BlochEigen}. Upon rescaling \eqref{Scaling}, we consider a finite number $(2 N_{\max} + 1)^2$ of reciprocal lattice vectors $\bm G$, ${\bm G}'$ where the integer $N_{\max}>0$ represents the `maximum plane wave number' \citep{barnwell16}. This step amounts to a truncation of spatial Fourier series at order $N_{\max}$. Next, the $(2 N_{\max} + 1)^2 \times (2 N_{\max} + 1)^2$ matrices $\bf M$, $\bf K$ of Eq.~\eqref{BlochEigen} are constructed. In practice, the integrals \eqref{FourierCoeffs} defining $\bm{\mathcal{A}}$ and $\bm{\mathcal{R}}$ are computed numerically using left Riemann sums with $40\, N_{\max}$ points in each spatial direction{\,---\,}Riemann sums are equivalent to the trapezoidal rule since the integrand is periodic. The sum is evaluated using Matlab's Fast Fourier Transform algorithm \texttt{fft2}. At order zero in $g$, the elastic case is recovered, and the corresponding modal frequencies are computed using Matlab's \texttt{eig} function. Computations involved in the implementation of the perturbation method \eqref{PerturbApprox} are of the same nature as in the elastic case, thus following similar steps. 

Here, we used $N_{\max} = 6$ Fourier modes. We restricted the study to real-valued wavevectors such that $\text{Im}\,\check{\bm \kappa} = \bm{0}$ by setting $\bm{\kappa}^1 = \bm{0}$ for the viscoelastic perturbations, see Eq.~\eqref{PertMatrix}. In a standard fashion, normalised wavevectors $\check{\bm \kappa}$ scan the edges of the irreducible Brillouin zone\footnote{The present coordinates of M, $\Gamma$, X are used in related literature. Similarly to \citet{deymier2013acoustic} p.~97, the basis vectors for $\check{\bm \kappa}$ in \citet{barnwell16} should be understood `in units of $2\pi$', that is ${\bf i} = (2\pi,0)$, ${\bf j} = (0,2\pi)$. Note the typo in Fig.~2 of \citet{depascalis2020}.}, i.e. $\check{\bm \kappa}$ varies linearly along the path $\text{M:} (\pi,\pi) \to \Gamma\text{:}(0,0) \to \text{X:}(\pi,0) \to \text{M:}(\pi,\pi)$. Here, we have set 20 points along each edge, and results are obtained in a reasonable computational time.

\begin{figure}
	\centering
	\includegraphics[scale=0.40]{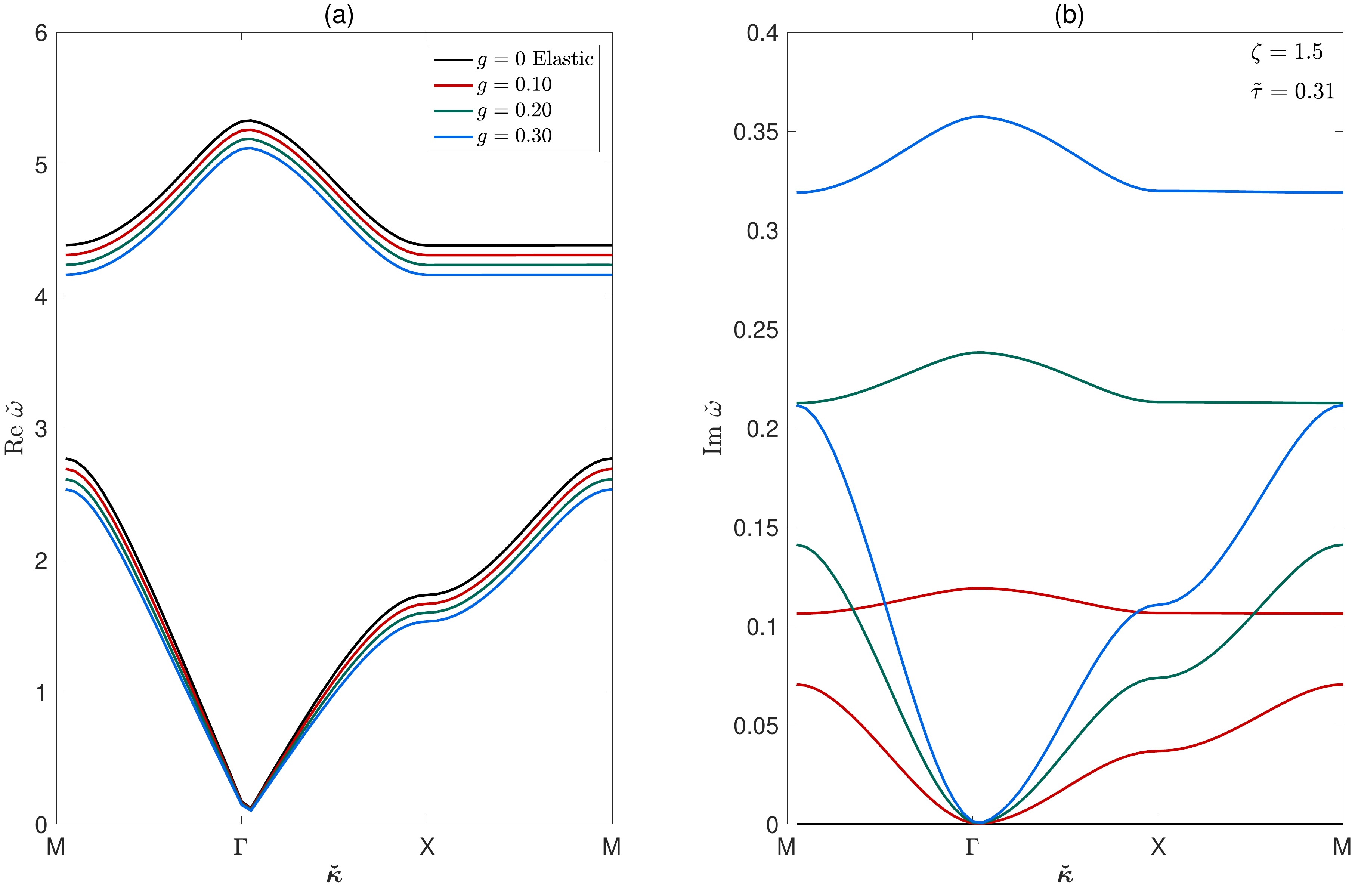}
	\caption{Dispersion curves in the phononic crystal for several values of $g$ while $\check\tau =0.31$ is kept constant. The stretch $\zeta = 1.5$ is applied to the cylinders, whose stiffness equals $10\%$ of the host material's stiffness given in Table~\ref{tab:Params}. (a) Real and (b) imaginary frequency.} \label{fig:dispersion_curves_g}
\end{figure}

Fig.~\ref{fig:dispersion_curves_g} shows the band diagram for the real and imaginary part of $\check{\omega}$, where each curve corresponds to a given propagation mode{\,---\,}here the first two modes are shown. Those curves refer to the pre-stressed case $\zeta=1.5$ while $g$ varies as shown in the legend. The blank space between consecutive real frequencies corresponds to the first \emph{band gap} (Fig.~\ref{fig:dispersion_curves_g}a). Remembering that $g=0$ represents the elastic case (black solid line), we can observe that the displayed modes shift towards lower frequencies as $g$ is increased (with a roughly linear  dependence). Since the imaginary part of $\check{\omega}$ is non-negative, a dissipative dynamic behaviour is found (Fig.~\ref{fig:dispersion_curves_g}b). The figure shows that the temporal attenuation increases with $g$.

\begin{figure}
	\centering
	\includegraphics[scale=0.40]{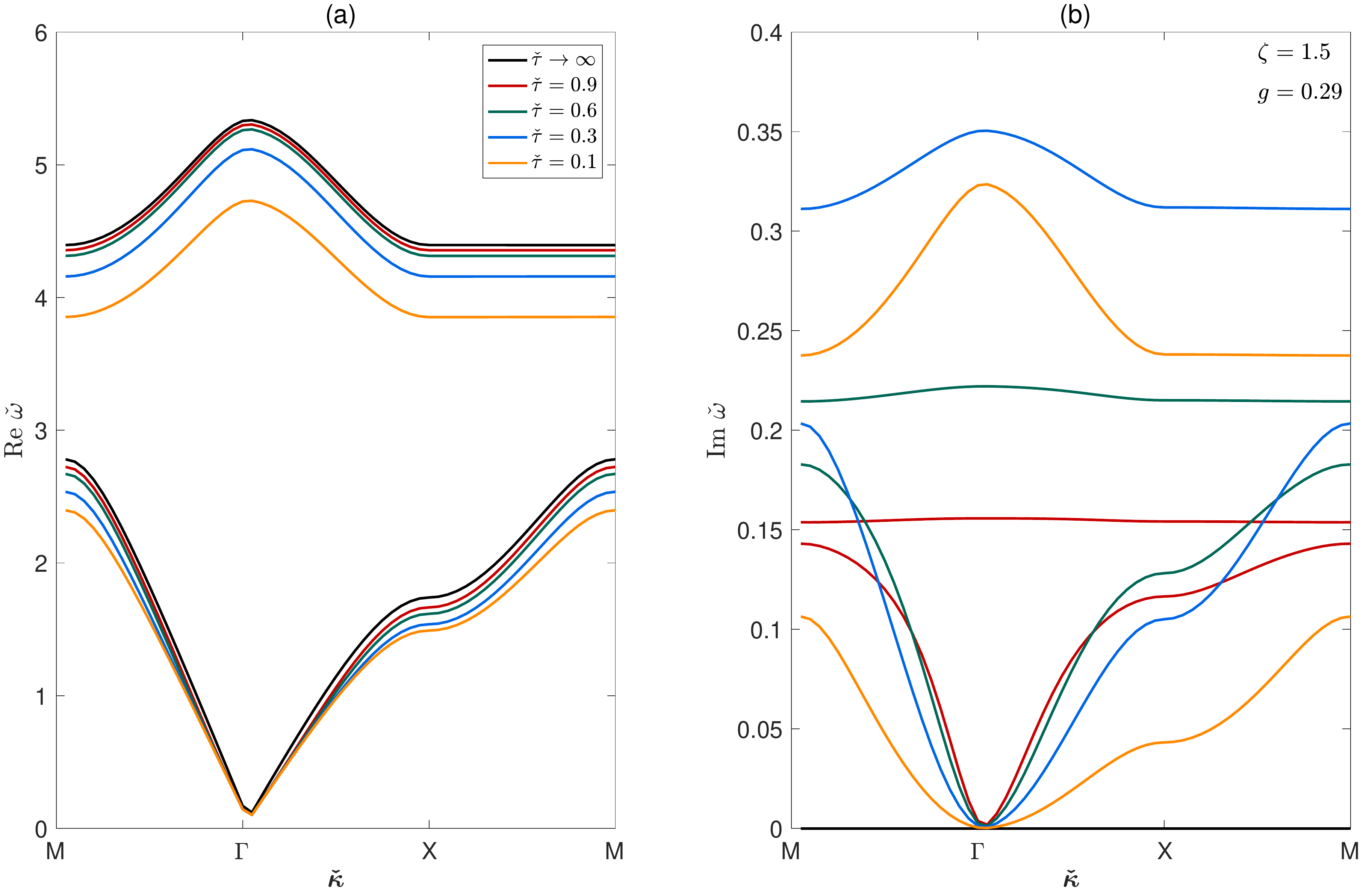}
	\caption{Dispersion curves in the phononic crystal for several values of $\check\tau$ while $g=0.29$ is kept constant. Same configuration as Fig.~\ref{fig:dispersion_curves_g}. \label{fig:dispersion_curves_tau}}
\end{figure}

Fig. \ref{fig:dispersion_curves_tau} is similar to Fig.~\ref{fig:dispersion_curves_g}, but now $\check{\tau}$ is varied while the other parameters are kept fixed (therefore $g=0.29$ and $\zeta=1.5$). We observe that Re $\check{\omega}$ is shifted nonlinearly towards higher frequencies as $\check{\tau}$ is increased. The limit $\check{\tau}\to +\infty$ corresponds to an elastic response, for which Im $\check{\omega} \to 0$. However, it is worth noting that this limit differs from the limit $g\to 0$, as shown in Eq.~\eqref{CoeffsBarnwell} and related discussions. The temporal attenuation in Fig.~\ref{fig:dispersion_curves_tau}b is evolving in a non-monotonous way with respect to $\check\tau$, a phenomenon that will be discussed later on. The same band diagrams were also produced at $\zeta=1$ (no pre-stress) and $\zeta=3$, leading to similar effects as in Figs.~\ref{fig:dispersion_curves_g}-\ref{fig:dispersion_curves_tau} upon varying $g$ and $\check{\tau}$.

\begin{figure}
	\centering
	\includegraphics[scale=0.40]{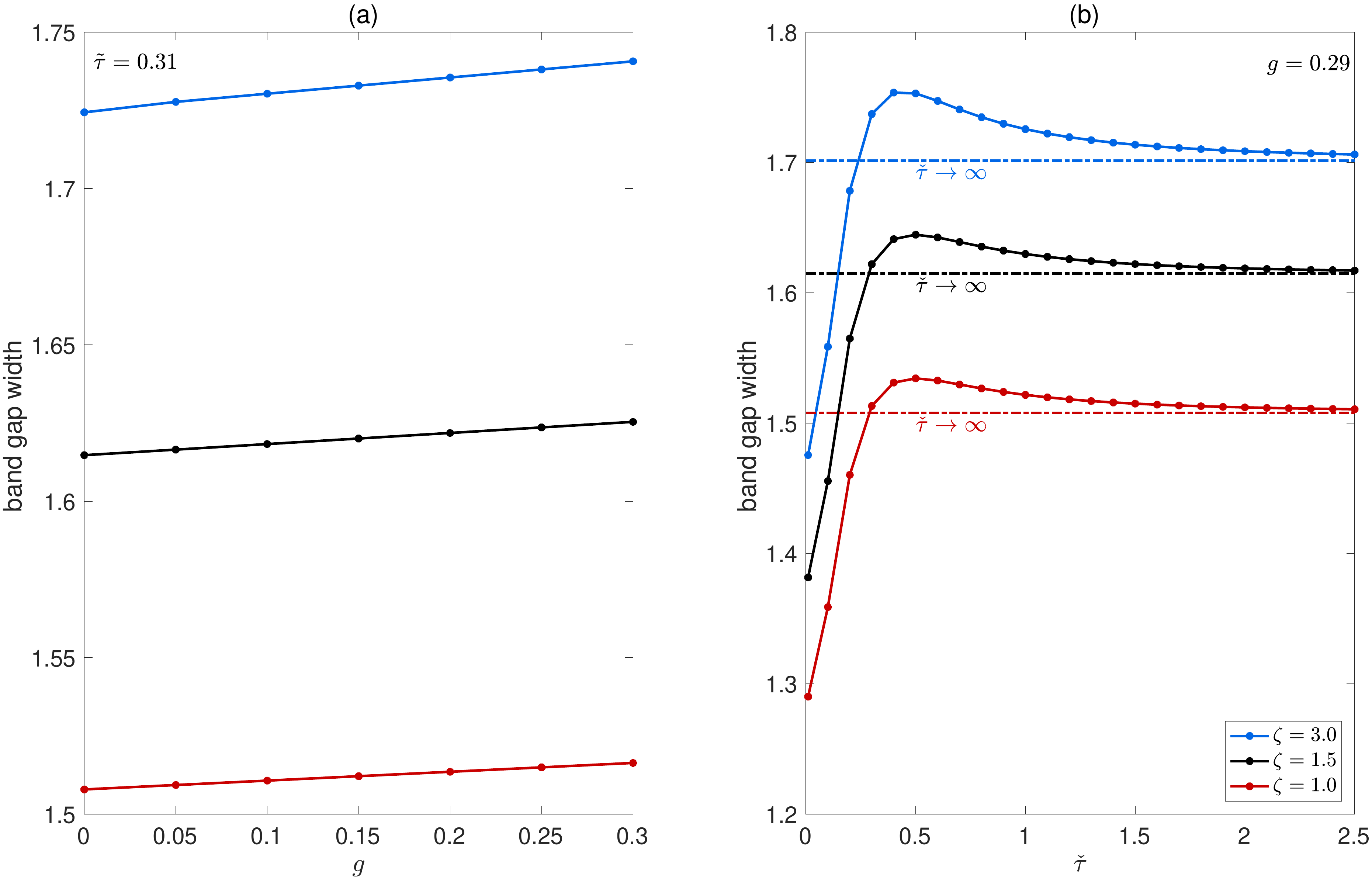}
	\caption{Band gap width referring to Figs.~\ref{fig:dispersion_curves_g}a-\ref{fig:dispersion_curves_tau}a (black solid line) in terms of $g$ or $\check\tau$ with other parameters kept constant. Other curves display the band gap width obtained for distinct stretching levels. \label{fig:bg_width}}
\end{figure}

The estimation of the band gap width is of particular interest for wave filtering once loss is taken into account ($g>0$). This is measured versus both $g$ and $\check{\tau}$ when all the other parameters are kept constant as above. The results are summarized in Fig.~\ref{fig:bg_width}. As shown in Fig.~\ref{fig:bg_width}a, the band gap width increases linearly with the parameter $g$ in all three cases ($\zeta=1$, $\zeta=1.5$ and $\zeta=3$ as displayed in a common legend). A non-monotonous evolution is instead observed for the band gap width with respect to $\check{\tau}$. This nonlinear evolution is all the more marked as the level of applied pre-deformation is large.

\begin{figure}
	\centering
	\includegraphics[scale=0.40]{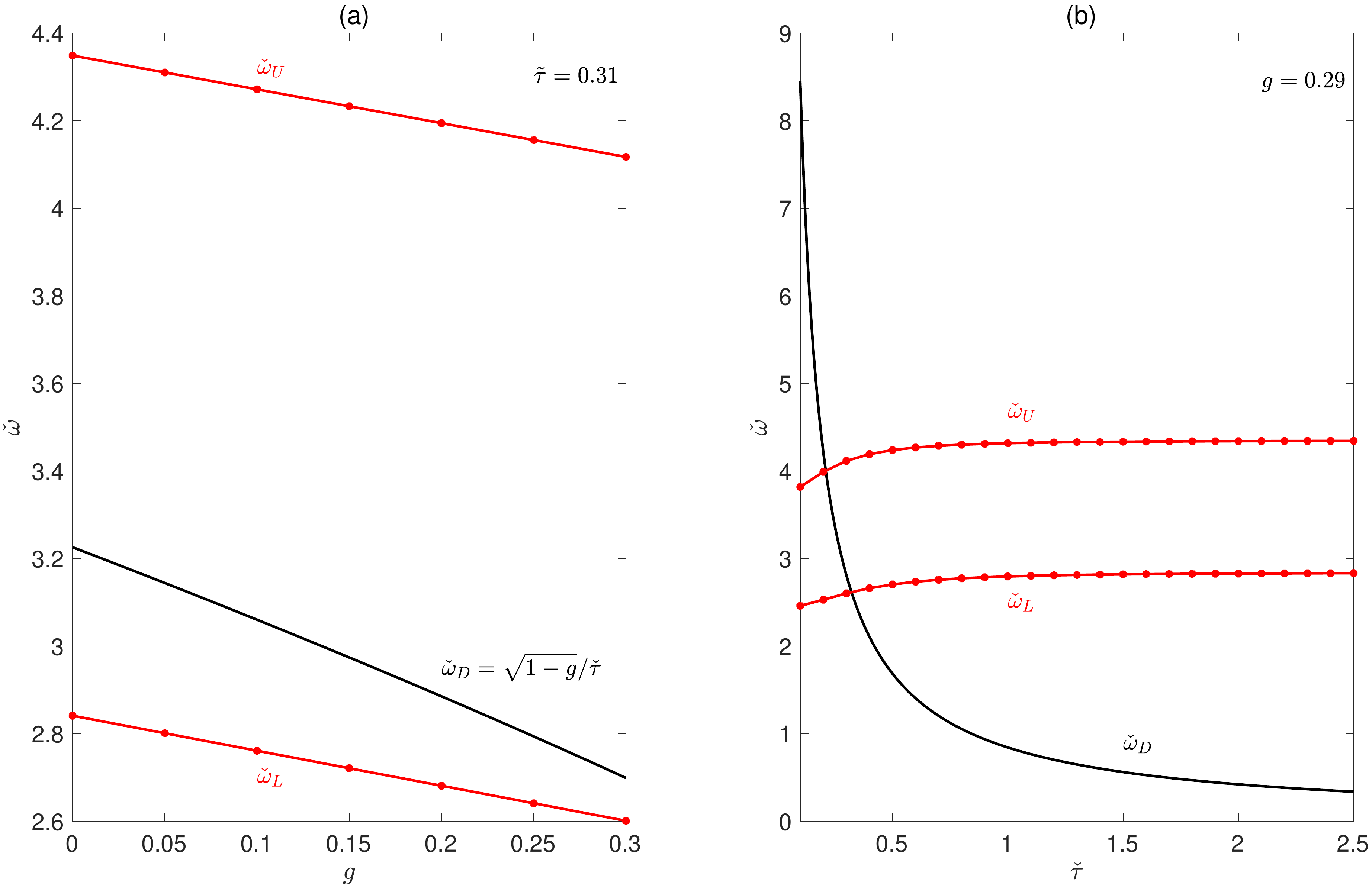}
	\caption{Frequency of maximum dissipation $\check\omega_D$ in the homogeneous solid, and band gap frequencies $\check\omega_L < \check\omega_U$ in the undeformed phononic structure $\zeta = 1$ in terms of (a) the parameter $g$ and (b) the parameter $\check\tau$. \label{fig:dissipation_unstressed}}
\end{figure}

For an unstressed homogeneous medium, the maximum dissipation $D=g/(2 \sqrt{1-g})$ is obtained at the scaled frequency $\check{\omega}_D=\sqrt{1-g}/\check{\tau}$  (see Sec.~\ref{sec:IncrDisp}). Clearly, as shown in Fig.~\ref{fig:dissipation_unstressed}a, $\check{\omega}_D$ remains within the band gap bounds $\check{\omega}_L < \check{\omega}_U$ when $g$ is varied (we refer to $\check{\omega}_L$ as the maximum frequency attained along the first mode, while $\check{\omega}_U$ refers to the minimum frequency attained along the second mode). However, at the same time, the level of dissipation $D$ is increasing with increasing values of $g$. This observation explains qualitatively why the band gap width is increasing with $g$ (Fig.~\ref{fig:bg_width}a), leading to the formulation of the following empirical conjecture:
\begin{quote}
	{\bfseries Conjecture.} \emph{With the present configuration, the more dissipation occurs in a band gap, the larger its width.}
\end{quote}

Now, let us look at the non-monotonous evolution of the attenuation and of the band gap width with respect to $\check\tau$ (Figs.~\ref{fig:dispersion_curves_tau}b-\ref{fig:bg_width}b). As shown in Fig.~\ref{fig:dissipation_unstressed}b, maximum dissipation occurs approximately within the range $0.1\leq\check\tau\leq 0.3$. Coherently, Fig.~\ref{fig:dispersion_curves_tau}b indicates that attenuation is very large in this range. However, Fig.~\ref{fig:bg_width}b shows that the band gaps are largest around $\check\tau \approx 0.5$. Therefore maximum dissipation in the homogeneous solid does not exactly entail the largest band gaps in the phononic crystal. Possibly this mismatch is due to the heterogeneity of the periodic structure.

In the end, it seems that the above empirical conjecture gives only a qualitative explanation for the non-monotonous evolution of the band gap width. In fact, if $\check\tau$ is very small, then $\check{\omega}_D$ is very large and the material is nearly elastic: almost no dissipation takes place in the frequency range $\check{\omega}_L < \check\omega < \check{\omega}_U$ of the first few propagation modes. In contrast, if $\check\tau$ is very large, then $\check{\omega}_D$ is very small; again, the material is nearly elastic in this frequency range of interest (with a different elastic limit than for $\check\tau \to 0$). Finally, viscoelastic dissipation really influences the band gap width when $\check\tau$ is neither too large nor too small. The above observations are reminiscent of \citet{zhao09} who found that ``the viscoelastic constants of host material affect not only the location but also the width of band gaps.''

For the aim of band gap tuning, the influence of elastic and geometric parameters was already discussed in the literature \citep{depascalis2020}. Above observations highlight the influence of viscoelastic parameters when one single relaxation mechanism is considered. On the one hand, we note that the band gap width increases with increasing values of $g$. On the other hand, the band gap width increases with increasing values of $\check\tau$ from $\check\tau \simeq 0$ up to a local maximum. If $\check\tau \to +\infty$ is increased further, then the band gap width decreases slowly towards a horizontal asymptote, i.e. the band gap width becomes less sensitive to variations of $\check\tau$. While the viscoelastic constitutive parameters cannot be adjusted easily in practice, we emphasize that real elastomers are better described by a sequence of relaxation mechanisms \citep{ciambella10}. Therefore, one of the numerous relaxation times $\check\tau$ might be large enough for the corresponding relaxation mechanism to contribute to the increase of the band gap width.

\section{Conclusion}\label{sec:Conclusion}

Phononic crystals are manufactured materials designed for the purpose of controlling sound and vibration based on their tunable geometric and material properties. In the last two decades, numerous research works have focussed on these materials and their applications in engineering, from electronic devices such as diodes and transistors to noise control devices \citep{khelif2015phononic}. Despite increasing interest, relevant literature on lossy materials subject to large deformations is rather scarce, to  the authors' present knowledge.

In summary, the present study addresses (i) the computation of incremental stresses in incompressible Fung--Simo solids; (ii) the analysis of dispersion for infinitesimal waves superimposed on a large static deformation; (iii) the Bloch-wave analysis for a lossy phononic crystal made of pre-stressed cylinders whose istantaneous elastic response is described by a Mooney--Rivlin potential. Key elements are the use of stress-like memory variables governed by linear evolution equations, as well as the implementation of a viscoelastic perturbation method.

Given that the present study is quasi-analytical, its full range of application remains quite restricted (small $g$, uniform dissipation, low frequency, propagative modes, one relaxation mechanism, etc.). Since computational approaches are more versatile, they seem to be a relevant tool for future works in this direction. Global wave attenuation properties are described in \citet{krushynska16} where the authors report ``an increase [of the wave attenuation performance] outside the band gap in the same way as in homogeneous materials''{\,---\,}a similar observation is reported in \citet{wang15} (not shown here).

Application to other materials could be considered, for instance viscoelastic materials of differential type \citep{destrade09}, compressible solids, phononic crystal plates \citep{mazzotti19}, and other systems involving slender structures \citep{amendola18}. Further developments could also encompass the study of pre-deformed composites \citep{huang14,Galich-Rudykh2017}, topological insulators \citep{nguyen19}, as well as periodic media based on electro-active or magneto-active materials \citep{getz17,KaramiMohammadi19}. In this context, high-order dynamic homogenisation theories could provide effective models for pre-stressed viscoelastic structures valid at moderate frequencies \citep{hu17}. Another potential direction of research is the study of manufactured phononic crystals with an irregular lattice \citep{mukhopadhyay19}.

\section*{Acknowledgments}

The authors are grateful to William J. Parnell (University of Manchester) and Michel Destrade (NUI Galway) for fruitful discussions, careful reading and support. HB was supported by the Irish Research Council [project ID GOIPD/2019/328]. RDP was supported by  Regione Puglia (Italy) through the research programme `Research for Innovation' (REFIN) [project number UNISAL023 - protocol code 2BDDFA20] and partially supported  by Italian National Group for Mathematical Physics (GNFM-INdAM).

%Research for Innovation (REFIN) - POR PUGLIA FESR-FSE 2014/2020 project UNISAL023 with code 2BDDFA20.
%
%
%The author is supported by Regione Puglia (Italy) through the research programme `Research for In-
%novation' (REFIN) - POR PUGLIA FESR-FSE 2014/2020 (project number UNISAL023, protocl code
%2BDDFA20).

%%%%%%%%%%%%% References biblio %%%%%%%%%%%%%%%
%\addcontentsline{toc}{section}{References}
%
%%%------- HOMEMADE BIBLIO
%\printbibliography

%%%------------ MRC BIBLIO
%\section*{\refname}
%\addcontentsline{toc}{section}{References}
\bibliography{biblio7}{}

\begin{thebibliography}{46}
\expandafter\ifx\csname natexlab\endcsname\relax\def\natexlab#1{#1}\fi
\providecommand{\url}[1]{\texttt{#1}}
\providecommand{\href}[2]{#2}
\providecommand{\path}[1]{#1}
\providecommand{\DOIprefix}{doi:}
\providecommand{\ArXivprefix}{arXiv:}
\providecommand{\URLprefix}{URL: }
\providecommand{\Pubmedprefix}{pmid:}
\providecommand{\doi}[1]{\href{http://dx.doi.org/#1}{\path{#1}}}
\providecommand{\Pubmed}[1]{\href{pmid:#1}{\path{#1}}}
\providecommand{\bibinfo}[2]{#2}
\ifx\xfnm\relax \def\xfnm[#1]{\unskip,\space#1}\fi
%Type = Book
\bibitem[{Al~Mayah(2018)}]{al2018biomechanics}
\bibinfo{author}{Al~Mayah, A.}, \bibinfo{year}{2018}.
\newblock \bibinfo{title}{Biomechanics of Soft Tissues}.
\newblock \bibinfo{publisher}{CRC Press}.
\newblock \DOIprefix\doi{10.1201/9781351135825}.
%Type = Article
\bibitem[{Amendola et~al.(2018)Amendola, Krushynska, Daraio, Pugno and
  Fraternali}]{amendola18}
\bibinfo{author}{Amendola, A.}, \bibinfo{author}{Krushynska, A.},
  \bibinfo{author}{Daraio, C.}, \bibinfo{author}{Pugno, N.M.},
  \bibinfo{author}{Fraternali, F.}, \bibinfo{year}{2018}.
\newblock \bibinfo{title}{Tuning frequency band gaps of tensegrity mass-spring
  chains with local and global prestress}.
\newblock \bibinfo{journal}{Int. J. Solids Struct.} \bibinfo{volume}{155},
  \bibinfo{pages}{47--56}.
\newblock \DOIprefix\doi{10.1016/j.ijsolstr.2018.07.002}.
%Type = Article
\bibitem[{Andreassen and Jensen(2013)}]{andreassen2013}
\bibinfo{author}{Andreassen, E.}, \bibinfo{author}{Jensen, J.S.},
  \bibinfo{year}{2013}.
\newblock \bibinfo{title}{Analysis of phononic bandgap structures with
  dissipation}.
\newblock \bibinfo{journal}{J. Vib. Acoust.} \bibinfo{volume}{135},
  \bibinfo{pages}{041015}.
\newblock \DOIprefix\doi{10.1115/1.4023901}.
%Type = Article
\bibitem[{Balbi et~al.(2018)Balbi, Shearer and Parnell}]{balbi18}
\bibinfo{author}{Balbi, V.}, \bibinfo{author}{Shearer, T.},
  \bibinfo{author}{Parnell, W.J.}, \bibinfo{year}{2018}.
\newblock \bibinfo{title}{A modified formulation of quasi-linear
  viscoelasticity for transversely isotropic materials under finite
  deformation}.
\newblock \bibinfo{journal}{Proc. R. Soc. A} \bibinfo{volume}{474},
  \bibinfo{pages}{20180231}.
\newblock \DOIprefix\doi{10.1098/rspa.2018.0231}.
%Type = Article
\bibitem[{Banks et~al.(2011)Banks, Hu and Kenz}]{banks11}
\bibinfo{author}{Banks, H.T.}, \bibinfo{author}{Hu, S.}, \bibinfo{author}{Kenz,
  Z.R.}, \bibinfo{year}{2011}.
\newblock \bibinfo{title}{A brief review of elasticity and viscoelasticity for
  solids}.
\newblock \bibinfo{journal}{Adv. Appl. Math. Mech.} \bibinfo{volume}{3},
  \bibinfo{pages}{1--51}.
\newblock \DOIprefix\doi{10.4208/aamm.10-m1030}.
%Type = Article
\bibitem[{Barnwell et~al.(2016)Barnwell, Parnell and Abrahams}]{barnwell16}
\bibinfo{author}{Barnwell, E.G.}, \bibinfo{author}{Parnell, W.J.},
  \bibinfo{author}{Abrahams, I.D.}, \bibinfo{year}{2016}.
\newblock \bibinfo{title}{Antiplane elastic wave propagation in pre-stressed
  periodic structures; tuning, band gap switching and invariance}.
\newblock \bibinfo{journal}{Wave Motion} \bibinfo{volume}{63},
  \bibinfo{pages}{98--110}.
\newblock \DOIprefix\doi{10.1016/j.wavemoti.2016.02.001}.
%Type = Article
\bibitem[{Berjamin et~al.(2021)Berjamin, Destrade and Parnell}]{berjamin20}
\bibinfo{author}{Berjamin, H.}, \bibinfo{author}{Destrade, M.},
  \bibinfo{author}{Parnell, W.J.}, \bibinfo{year}{2021}.
\newblock \bibinfo{title}{On the thermodynamic consistency of quasi-linear
  viscoelastic models for soft solids}.
\newblock \bibinfo{journal}{Mech. Res. Commun.} \bibinfo{volume}{111},
  \bibinfo{pages}{103648}.
\newblock \DOIprefix\doi{10.1016/j.mechrescom.2020.103648}.
%Type = Book
\bibitem[{Carcione(2015)}]{carcione15}
\bibinfo{author}{Carcione, J.M.}, \bibinfo{year}{2015}.
\newblock \bibinfo{title}{Wave Fields in Real Media}.
\newblock \bibinfo{edition}{3} ed., \bibinfo{publisher}{Elsevier},
  \bibinfo{address}{Amsterdam}.
\newblock \DOIprefix\doi{10.1016/C2013-0-18893-9}.
%Type = Article
\bibitem[{Ciambella et~al.(2010)Ciambella, Paolone and Vidoli}]{ciambella10}
\bibinfo{author}{Ciambella, J.}, \bibinfo{author}{Paolone, A.},
  \bibinfo{author}{Vidoli, S.}, \bibinfo{year}{2010}.
\newblock \bibinfo{title}{A comparison of nonlinear integral-based viscoelastic
  models through compression tests on filled rubber}.
\newblock \bibinfo{journal}{Mech. Mater.} \bibinfo{volume}{42},
  \bibinfo{pages}{932--944}.
\newblock \DOIprefix\doi{10.1016/j.mechmat.2010.07.007}.
%Type = Article
\bibitem[{De~Pascalis et~al.(2014)De~Pascalis, Abrahams and
  Parnell}]{depascalis14}
\bibinfo{author}{De~Pascalis, R.}, \bibinfo{author}{Abrahams, I.D.},
  \bibinfo{author}{Parnell, W.J.}, \bibinfo{year}{2014}.
\newblock \bibinfo{title}{On nonlinear viscoelastic deformations: a reappraisal
  of {F}ung's quasi-linear viscoelastic model}.
\newblock \bibinfo{journal}{Proc. R. Soc. A} \bibinfo{volume}{470},
  \bibinfo{pages}{20140058}.
\newblock \DOIprefix\doi{10.1098/rspa.2014.0058}.
%Type = Article
\bibitem[{De~Pascalis et~al.(2011)De~Pascalis, Destrade and Goriely}]{dpr11}
\bibinfo{author}{De~Pascalis, R.}, \bibinfo{author}{Destrade, M.},
  \bibinfo{author}{Goriely, A.}, \bibinfo{year}{2011}.
\newblock \bibinfo{title}{Nonlinear correction to the {E}uler buckling formula
  for compressed cylinders with guided-guided end conditions}.
\newblock \bibinfo{journal}{J. Elast.} \bibinfo{volume}{102},
  \bibinfo{pages}{191--200}.
\newblock \DOIprefix\doi{10.1007/s10659-010-9265-6}.
%Type = Article
\bibitem[{De~Pascalis et~al.(2020)De~Pascalis, Donateo, Ficarella and
  Parnell}]{depascalis2020}
\bibinfo{author}{De~Pascalis, R.}, \bibinfo{author}{Donateo, T.},
  \bibinfo{author}{Ficarella, A.}, \bibinfo{author}{Parnell, W.J.},
  \bibinfo{year}{2020}.
\newblock \bibinfo{title}{Optimal design of phononic media through genetic
  algorithm-informed pre-stress for the control of antiplane wave propagation}.
\newblock \bibinfo{journal}{Extreme Mech. Lett.} \bibinfo{volume}{40},
  \bibinfo{pages}{100896}.
\newblock \DOIprefix\doi{10.1016/j.eml.2020.100896}.
%Type = Article
\bibitem[{De~Pascalis et~al.(2018)De~Pascalis, Parnell, Abrahams, Shearer, Daly
  and Grundy}]{dpr18b}
\bibinfo{author}{De~Pascalis, R.}, \bibinfo{author}{Parnell, W.J.},
  \bibinfo{author}{Abrahams, I.D.}, \bibinfo{author}{Shearer, T.},
  \bibinfo{author}{Daly, D.M.}, \bibinfo{author}{Grundy, D.},
  \bibinfo{year}{2018}.
\newblock \bibinfo{title}{The inflation of viscoelastic balloons and hollow
  viscera}.
\newblock \bibinfo{journal}{Proc. R. Soc. A} \bibinfo{volume}{474},
  \bibinfo{pages}{20180102}.
\newblock \DOIprefix\doi{10.1098/rspa.2018.0102}.
%Type = Article
\bibitem[{Destrade et~al.(2009)Destrade, Ogden and Saccomandi}]{destrade09}
\bibinfo{author}{Destrade, M.}, \bibinfo{author}{Ogden, R.W.},
  \bibinfo{author}{Saccomandi, G.}, \bibinfo{year}{2009}.
\newblock \bibinfo{title}{Small amplitude waves and stability for a
  pre-stressed viscoelastic solid}.
\newblock \bibinfo{journal}{Z. angew. Math. Phys.} \bibinfo{volume}{60},
  \bibinfo{pages}{511--528}.
\newblock \DOIprefix\doi{10.1007/s00033-008-7147-6}.
%Type = Book
\bibitem[{Deymier(2013)}]{deymier2013acoustic}
\bibinfo{author}{Deymier, P.A.}, \bibinfo{year}{2013}.
\newblock \bibinfo{title}{Acoustic Metamaterials and Phononic Crystals}.
\newblock \bibinfo{publisher}{Springer}, \bibinfo{address}{Berlin, Heidelberg}.
\newblock \DOIprefix\doi{10.1007/978-3-642-31232-8}.
%Type = Book
\bibitem[{Fung(1993)}]{fung93}
\bibinfo{author}{Fung, Y.C.}, \bibinfo{year}{1993}.
\newblock \bibinfo{title}{Biomechanics}.
\newblock \bibinfo{edition}{2} ed., \bibinfo{publisher}{Springer-Verlag},
  \bibinfo{address}{New York}.
\newblock \DOIprefix\doi{10.1007/978-1-4757-2257-4}.
%Type = Article
\bibitem[{Galich et~al.(2017)Galich, Slesarenko and Rudykh}]{Galich-Rudykh2017}
\bibinfo{author}{Galich, P.I.}, \bibinfo{author}{Slesarenko, V.},
  \bibinfo{author}{Rudykh, S.}, \bibinfo{year}{2017}.
\newblock \bibinfo{title}{Shear wave propagation in finitely deformed 3{D}
  fiber-reinforced composites}.
\newblock \bibinfo{journal}{Int. J. Solids Struct.} \bibinfo{volume}{110-111},
  \bibinfo{pages}{294--304}.
\newblock \DOIprefix\doi{10.1016/j.ijsolstr.2016.12.007}.
%Type = Article
\bibitem[{Getz et~al.(2017)Getz, Kochmann and Shmuel}]{getz17}
\bibinfo{author}{Getz, R.}, \bibinfo{author}{Kochmann, D.M.},
  \bibinfo{author}{Shmuel, G.}, \bibinfo{year}{2017}.
\newblock \bibinfo{title}{Voltage-controlled complete stopbands in
  two-dimensional soft dielectrics}.
\newblock \bibinfo{journal}{Int. J. Solids Struct.} \bibinfo{volume}{113},
  \bibinfo{pages}{24--36}.
\newblock \DOIprefix\doi{10.1016/j.ijsolstr.2016.10.002}.
%Type = Article
\bibitem[{Goriely et~al.(2008)Goriely, Vandiver and
  Destrade}]{goriely_vandiver_destrade2008}
\bibinfo{author}{Goriely, A.}, \bibinfo{author}{Vandiver, R.},
  \bibinfo{author}{Destrade, M.}, \bibinfo{year}{2008}.
\newblock \bibinfo{title}{Nonlinear {E}uler buckling}.
\newblock \bibinfo{journal}{Proc. R. Soc. A} \bibinfo{volume}{464},
  \bibinfo{pages}{3003--3019}.
\newblock \DOIprefix\doi{10.1098/rspa.2008.0184}.
%Type = Article
\bibitem[{Helisaz et~al.(2021)Helisaz, Bacca and Chiao}]{helisaz2021}
\bibinfo{author}{Helisaz, H.}, \bibinfo{author}{Bacca, M.},
  \bibinfo{author}{Chiao, M.}, \bibinfo{year}{2021}.
\newblock \bibinfo{title}{Quasi-linear viscoelastic characterization of soft
  tissue-mimicking materials}.
\newblock \bibinfo{journal}{J. Biomech. Eng.} \bibinfo{volume}{143},
  \bibinfo{pages}{061007}.
\newblock \DOIprefix\doi{10.1115/1.4050036}.
%Type = Book
\bibitem[{Holzapfel(2000)}]{holzapfel00}
\bibinfo{author}{Holzapfel, G.A.}, \bibinfo{year}{2000}.
\newblock \bibinfo{title}{Nonlinear Solid Mechanics: A Continuum Approach for
  Engineering}.
\newblock \bibinfo{publisher}{John Wiley {\&} Sons Ltd.},
  \bibinfo{address}{Chichester}.
%Type = Book
\bibitem[{Holzapfel and Ogden(2003)}]{holzapfel2014biomechanics}
\bibinfo{author}{Holzapfel, G.A.}, \bibinfo{author}{Ogden, R.W.},
  \bibinfo{year}{2003}.
\newblock \bibinfo{title}{Biomechanics of Soft Tissue in Cardiovascular
  Systems}.
\newblock \bibinfo{publisher}{Springer}, \bibinfo{address}{Vienna}.
\newblock \DOIprefix\doi{10.1007/978-3-7091-2736-0}.
%Type = Article
\bibitem[{Hosler et~al.(1999)Hosler, Burkett and Tarkanian}]{hosler99}
\bibinfo{author}{Hosler, D.}, \bibinfo{author}{Burkett, S.L.},
  \bibinfo{author}{Tarkanian, M.J.}, \bibinfo{year}{1999}.
\newblock \bibinfo{title}{Prehistoric polymers: {R}ubber processing in ancient
  {M}esoamerica}.
\newblock \bibinfo{journal}{Science} \bibinfo{volume}{284},
  \bibinfo{pages}{1988--1991}.
\newblock \DOIprefix\doi{10.1126/science.284.5422.1988}.
%Type = Article
\bibitem[{Hu and Oskay(2017)}]{hu17}
\bibinfo{author}{Hu, R.}, \bibinfo{author}{Oskay, C.}, \bibinfo{year}{2017}.
\newblock \bibinfo{title}{Nonlocal homogenization model for wave dispersion and
  attenuation in elastic and viscoelastic periodic layered media}.
\newblock \bibinfo{journal}{J. Appl. Mech.} \bibinfo{volume}{84},
  \bibinfo{pages}{031003}.
\newblock \DOIprefix\doi{10.1115/1.4035364}.
%Type = Article
\bibitem[{Huang et~al.(2014)Huang, Shen, Zhang and Chen}]{huang14}
\bibinfo{author}{Huang, Y.}, \bibinfo{author}{Shen, X.D.},
  \bibinfo{author}{Zhang, C.L.}, \bibinfo{author}{Chen, W.Q.},
  \bibinfo{year}{2014}.
\newblock \bibinfo{title}{Mechanically tunable band gaps in compressible soft
  phononic laminated composites with finite deformation}.
\newblock \bibinfo{journal}{Phys. Lett. A} \bibinfo{volume}{378},
  \bibinfo{pages}{2285--2289}.
\newblock \DOIprefix\doi{10.1016/j.physleta.2014.05.032}.
%Type = Article
\bibitem[{Hussein(2009)}]{hussein2009}
\bibinfo{author}{Hussein, M.I.}, \bibinfo{year}{2009}.
\newblock \bibinfo{title}{Theory of damped {B}loch waves in elastic media}.
\newblock \bibinfo{journal}{Phys. Rev. B} \bibinfo{volume}{80},
  \bibinfo{pages}{212301}.
\newblock \DOIprefix\doi{10.1103/PhysRevB.80.212301}.
%Type = Article
\bibitem[{Hussein and Frazier(2010)}]{hussein2010}
\bibinfo{author}{Hussein, M.I.}, \bibinfo{author}{Frazier, M.J.},
  \bibinfo{year}{2010}.
\newblock \bibinfo{title}{Band structure of phononic crystals with general
  damping}.
\newblock \bibinfo{journal}{J. Appl. Phys.} \bibinfo{volume}{108},
  \bibinfo{pages}{093506}.
\newblock \DOIprefix\doi{10.1063/1.3498806}.
%Type = Article
\bibitem[{Jridi et~al.(2019)Jridi, Arfaoui, Hamdi, Salvia, Bareille, Ichchou
  and Ben~Abdallah}]{jridi19}
\bibinfo{author}{Jridi, N.}, \bibinfo{author}{Arfaoui, M.},
  \bibinfo{author}{Hamdi, A.}, \bibinfo{author}{Salvia, M.},
  \bibinfo{author}{Bareille, O.}, \bibinfo{author}{Ichchou, M.},
  \bibinfo{author}{Ben~Abdallah, J.}, \bibinfo{year}{2019}.
\newblock \bibinfo{title}{Separable finite viscoelasticity: integral-based
  models vs. experiments}.
\newblock \bibinfo{journal}{Mech. Time-Depend. Mater.} \bibinfo{volume}{23},
  \bibinfo{pages}{295--325}.
\newblock \DOIprefix\doi{10.1007/s11043-018-9383-2}.
%Type = Article
\bibitem[{Karami~Mohammadi et~al.(2019)Karami~Mohammadi, Galich, Krushynska and
  Rudykh}]{KaramiMohammadi19}
\bibinfo{author}{Karami~Mohammadi, N.}, \bibinfo{author}{Galich, P.I.},
  \bibinfo{author}{Krushynska, A.O.}, \bibinfo{author}{Rudykh, S.},
  \bibinfo{year}{2019}.
\newblock \bibinfo{title}{Soft magnetoactive laminates: large deformations,
  transverse elastic waves and band gaps tunability by a magnetic field}.
\newblock \bibinfo{journal}{J. Appl. Mech.} \bibinfo{volume}{86},
  \bibinfo{pages}{111001}.
\newblock \DOIprefix\doi{10.1115/1.4044497}.
%Type = Article
\bibitem[{Khajehsaeid et~al.(2013)Khajehsaeid, Arghavani and
  Naghdabadi}]{khajehsaeid13}
\bibinfo{author}{Khajehsaeid, H.}, \bibinfo{author}{Arghavani, J.},
  \bibinfo{author}{Naghdabadi, R.}, \bibinfo{year}{2013}.
\newblock \bibinfo{title}{A hyperelastic constitutive model for rubber-like
  materials}.
\newblock \bibinfo{journal}{Eur. J. Mech. A-Solids} \bibinfo{volume}{38},
  \bibinfo{pages}{144--151}.
\newblock \DOIprefix\doi{10.1016/j.euromechsol.2012.09.010}.
%Type = Book
\bibitem[{Khelif and Adibi(2016)}]{khelif2015phononic}
\bibinfo{author}{Khelif, A.}, \bibinfo{author}{Adibi, A.},
  \bibinfo{year}{2016}.
\newblock \bibinfo{title}{Phononic Crystals}.
\newblock \bibinfo{publisher}{Springer-Verlag}, \bibinfo{address}{New York}.
\newblock \DOIprefix\doi{10.1007/978-1-4614-9393-8}.
%Type = Article
\bibitem[{Krushynska et~al.(2016)Krushynska, Kouznetsova and
  Geers}]{krushynska16}
\bibinfo{author}{Krushynska, A.O.}, \bibinfo{author}{Kouznetsova, V.G.},
  \bibinfo{author}{Geers, M.G.D.}, \bibinfo{year}{2016}.
\newblock \bibinfo{title}{Visco-elastic effects on wave dispersion in
  three-phase acoustic metamaterials}.
\newblock \bibinfo{journal}{J. Mech. Phys. Solids} \bibinfo{volume}{96},
  \bibinfo{pages}{29–47}.
\newblock \DOIprefix\doi{10.1016/j.jmps.2016.07.001}.
%Type = Article
\bibitem[{Li et~al.(2021)Li, Yang, Ma and Xia}]{li21}
\bibinfo{author}{Li, J.}, \bibinfo{author}{Yang, P.}, \bibinfo{author}{Ma, Q.},
  \bibinfo{author}{Xia, M.}, \bibinfo{year}{2021}.
\newblock \bibinfo{title}{Complex band structure and attenuation performance of
  a viscoelastic phononic crystal with finite out-of-plane extension}.
\newblock \bibinfo{journal}{Acta Mech.} \bibinfo{volume}{232},
  \bibinfo{pages}{2933–2954}.
\newblock \DOIprefix\doi{10.1007/s00707-021-02969-8}.
%Type = Article
\bibitem[{Marckmann and Verron(2006)}]{marckmann06}
\bibinfo{author}{Marckmann, G.}, \bibinfo{author}{Verron, E.},
  \bibinfo{year}{2006}.
\newblock \bibinfo{title}{Comparison of hyperelastic models for rubber-like
  materials}.
\newblock \bibinfo{journal}{Rubber Chem. Technol.} \bibinfo{volume}{79},
  \bibinfo{pages}{835--858}.
\newblock \DOIprefix\doi{10.5254/1.3547969}.
%Type = Article
\bibitem[{Mazzotti et~al.(2019)Mazzotti, Bartoli and Miniaci}]{mazzotti19}
\bibinfo{author}{Mazzotti, M.}, \bibinfo{author}{Bartoli, I.},
  \bibinfo{author}{Miniaci, M.}, \bibinfo{year}{2019}.
\newblock \bibinfo{title}{Modeling {B}loch waves in prestressed phononic
  crystal plates}.
\newblock \bibinfo{journal}{Front. Mater.} \bibinfo{volume}{6},
  \bibinfo{pages}{74}.
\newblock \DOIprefix\doi{10.3389/fmats.2019.00074}.
%Type = Article
\bibitem[{Mokhtari et~al.(2019)Mokhtari, Lu and Srivastava}]{mokhtari19}
\bibinfo{author}{Mokhtari, A.A.}, \bibinfo{author}{Lu, Y.},
  \bibinfo{author}{Srivastava, A.}, \bibinfo{year}{2019}.
\newblock \bibinfo{title}{On the properties of phononic eigenvalue problems}.
\newblock \bibinfo{journal}{J. Mech. Phys. Solids} \bibinfo{volume}{131},
  \bibinfo{pages}{167--179}.
\newblock \DOIprefix\doi{10.1016/j.jmps.2019.07.005}.
%Type = Article
\bibitem[{Mukhopadhyay et~al.(2019)Mukhopadhyay, Adhikari and
  Batou}]{mukhopadhyay19}
\bibinfo{author}{Mukhopadhyay, T.}, \bibinfo{author}{Adhikari, S.},
  \bibinfo{author}{Batou, A.}, \bibinfo{year}{2019}.
\newblock \bibinfo{title}{Frequency domain homogenization for the viscoelastic
  properties of spatially correlated quasi-periodic lattices}.
\newblock \bibinfo{journal}{Int. J. Mech. Sci.} \bibinfo{volume}{150},
  \bibinfo{pages}{784--806}.
\newblock \DOIprefix\doi{10.1016/j.ijmecsci.2017.09.004}.
%Type = Article
\bibitem[{Nguyen et~al.(2019)Nguyen, Zhuang, Park and Rabczuk}]{nguyen19}
\bibinfo{author}{Nguyen, B.H.}, \bibinfo{author}{Zhuang, X.},
  \bibinfo{author}{Park, H.S.}, \bibinfo{author}{Rabczuk, T.},
  \bibinfo{year}{2019}.
\newblock \bibinfo{title}{Tunable topological bandgaps and frequencies in a
  pre-stressed soft phononic crystal}.
\newblock \bibinfo{journal}{J. Appl. Phys.} \bibinfo{volume}{125},
  \bibinfo{pages}{095106}.
\newblock \DOIprefix\doi{10.1063/1.5066088}.
%Type = Book
\bibitem[{Ogden(1984)}]{ogden1997non}
\bibinfo{author}{Ogden, R.W.}, \bibinfo{year}{1984}.
\newblock \bibinfo{title}{Non-Linear Elastic Deformations}.
\newblock \bibinfo{publisher}{Ellis Horwood Ltd.},
  \bibinfo{address}{Chichester}.
%Type = Article
\bibitem[{Parnell and De~Pascalis(2019)}]{parnell19}
\bibinfo{author}{Parnell, W.J.}, \bibinfo{author}{De~Pascalis, R.},
  \bibinfo{year}{2019}.
\newblock \bibinfo{title}{Soft metamaterials with dynamic viscoelastic
  functionality tuned by pre-deformation}.
\newblock \bibinfo{journal}{Phil. Trans. R. Soc. A} \bibinfo{volume}{377},
  \bibinfo{pages}{20180072}.
\newblock \DOIprefix\doi{10.1098/rsta.2018.0072}.
%Type = Article
\bibitem[{Scott and Hayes(1985)}]{scott85}
\bibinfo{author}{Scott, N.H.}, \bibinfo{author}{Hayes, M.},
  \bibinfo{year}{1985}.
\newblock \bibinfo{title}{A note on wave propagation in internally constrained
  hyperelastic materials}.
\newblock \bibinfo{journal}{Wave Motion} \bibinfo{volume}{7},
  \bibinfo{pages}{601--605}.
\newblock \DOIprefix\doi{10.1016/0165-2125(85)90037-X}.
%Type = Article
\bibitem[{Simo(1987)}]{simo87}
\bibinfo{author}{Simo, J.C.}, \bibinfo{year}{1987}.
\newblock \bibinfo{title}{On a fully three-dimensional finite-strain
  viscoelastic damage model: {F}ormulation and computational aspects}.
\newblock \bibinfo{journal}{Comput. Methods Appl. Mech. Engrg.}
  \bibinfo{volume}{60}, \bibinfo{pages}{153--173}.
\newblock \DOIprefix\doi{10.1016/0045-7825(87)90107-1}.
%Type = Article
\bibitem[{Taylor et~al.(2009)Taylor, Comas, Cheng, Passenger, Hawkes, Atkinson
  and Ourselin}]{taylor09}
\bibinfo{author}{Taylor, Z.A.}, \bibinfo{author}{Comas, O.},
  \bibinfo{author}{Cheng, M.}, \bibinfo{author}{Passenger, J.},
  \bibinfo{author}{Hawkes, D.J.}, \bibinfo{author}{Atkinson, D.},
  \bibinfo{author}{Ourselin, S.}, \bibinfo{year}{2009}.
\newblock \bibinfo{title}{On modelling of anisotropic viscoelasticity for soft
  tissue simulation: {N}umerical solution and {GPU} execution}.
\newblock \bibinfo{journal}{Med. Image Anal.} \bibinfo{volume}{13},
  \bibinfo{pages}{234--244}.
\newblock \DOIprefix\doi{10.1016/j.media.2008.10.001}.
%Type = Article
\bibitem[{Wang et~al.(2015)Wang, Wang and Laude}]{wang15}
\bibinfo{author}{Wang, Y.F.}, \bibinfo{author}{Wang, Y.S.},
  \bibinfo{author}{Laude, V.}, \bibinfo{year}{2015}.
\newblock \bibinfo{title}{Wave propagation in two-dimensional viscoelastic
  metamaterials}.
\newblock \bibinfo{journal}{Phys. Rev. B} \bibinfo{volume}{92},
  \bibinfo{pages}{104110}.
\newblock \DOIprefix\doi{10.1103/PhysRevB.92.104110}.
%Type = Article
\bibitem[{Wineman(2009)}]{wineman09}
\bibinfo{author}{Wineman, A.}, \bibinfo{year}{2009}.
\newblock \bibinfo{title}{Nonlinear viscoelastic solids{---}{A} review}.
\newblock \bibinfo{journal}{Math. Mech. Solids} \bibinfo{volume}{14},
  \bibinfo{pages}{300--366}.
\newblock \DOIprefix\doi{10.1177/1081286509103660}.
%Type = Article
\bibitem[{Zhao and Wei(2009)}]{zhao09}
\bibinfo{author}{Zhao, Y.P.}, \bibinfo{author}{Wei, P.J.},
  \bibinfo{year}{2009}.
\newblock \bibinfo{title}{The band gap of {1D} viscoelastic phononic crystal}.
\newblock \bibinfo{journal}{Comput. Mater. Sci.} \bibinfo{volume}{46},
  \bibinfo{pages}{603–606}.
\newblock \DOIprefix\doi{10.1016/j.commatsci.2009.03.040}.

\end{thebibliography}

%%%%%%%%%%%%% Appendix %%%%%%%%%%%%%%%
\appendix

\section{Thermodynamics}\label{app:thermo}

In this section, we present the non-equilibrium thermodynamic analysis of the present incompressible Fung--Simo theory by following the steps in \citet{berjamin20} (a compressible version of the Simo model is presented in the corresponding seminal works \citep{simo87}).
We introduce the thermodynamic potential
\begin{equation}
	\Psi = -q(J-1) + W(\tilde{\bm C}) - \tfrac12 \sum_{k=1}^n \left(\bm{S}_k^\text{v}:{\bm C} - \Phi_k(\bm{S}_k^\text{v}) \right)  ,
	\label{Energy}
\end{equation}
which is Helmholtz' free energy per unit of reference volume. The deformation tensor with overtilde is the volume-preserving strain tensor $\tilde{\bm C} = J^{-2/3}\bm{C}$. The functions $\Phi_k$ are presumably convex potentials to be determined. With the present definitions, the constitutive law \eqref{FungMemStress} reads $\bm{S} = 2\, \partial\Psi/\partial {\bm C}$ under the incompressibility constraint \eqref{Incomp}. If the pressure $p$ of Eq.~\eqref{FungConvSimo} was used instead of its redefinition $q$, then the tensor $\bm C$ in Eq.~\eqref{Energy} would have to be substituted by its volume-preserving version for consistency. Since the derivation is very similar with either expression, we restrict the presentation to the present one.

Assuming that the memory variables ${\bm S}_k^\text{v}$ governed by Eq.~\eqref{FungEvol} are \emph{internal variables} of state, the dissipation per unit volume is given by the corresponding formulas \citep{berjamin20}
\begin{equation}
	\mathscr{D} = - \sum_{k=1}^n \frac{\partial \Psi}{\partial \bm{S}_k^\text{v}} : \dot{\bm S}_k^\text{v} = \sum_{k=1}^n \frac{1}{2\tau_k} \left({\bm C} - \frac{\partial \Phi_k(\bm{S}_k^\text{v})}{\partial \bm{S}_k^\text{v}} \right) : \left(2g_k \frac{\partial W(\tilde{\bm C})}{\partial {\bm C}} - \bm{S}_k^\text{v}\right) .
	\label{Dissipation}
\end{equation}
We introduce the Legendre transform $2W_k = \bm{S}_k^\text{v}:{\bm C}_k^\text{v} - \Phi_k$ of the potential $\Phi_k$ such that the relationships ${\bm C}_k^\text{v} = \partial \Phi_k/\partial \bm{S}_k^\text{v}$ and $\bm{S}_k^\text{v} = 2\, \partial W_k/\partial {\bm C}_k^\text{v}$ are satisfied. As shown in \citet{berjamin20}, setting $W_k(\cdot) = g_k W(\tilde\cdot)$ then yields
\begin{equation}
	\mathscr{D} = \sum_{k=1}^n \frac{g_k}{\tau_k} \left( {\bm C} - {\bm C}_k^\text{v} \right) : \left(\frac{\partial W(\tilde{\bm C})}{\partial {\bm C}} - \frac{\partial W(\tilde{\bm C}_k^\text{v})}{\partial {\bm C}_k^\text{v}}\right) .
	\label{Legendre}
\end{equation}
The convexity inequality for $W$ entails the thermodynamic consistency of the Fung--Simo model.

\end{document}